\documentclass[fleqn,usenatbib]{mnras}

\usepackage{newtxtext,newtxmath}
\usepackage[]{amsmath}
\usepackage[T1]{fontenc}
\usepackage{ae,aecompl}

\usepackage{graphicx}	% Including figure files
\usepackage{amsmath}	% Advanced maths commands
\usepackage{amssymb}	% Extra maths symbols

\newcommand{\zobov}{{\small ZOBOV} }

\title[Accurate void-galaxy RSD]{An accurate linear model for redshift space distortions in the void-galaxy correlation function}
\author[S. Nadathur \& W. J. Percival]{Seshadri Nadathur$^1$\thanks{seshadri.nadathur@port.ac.uk} and Will J. Percival$^{2,3,1}$
\\
$^1$Institute of Cosmology and Gravitation, University of Portsmouth, Burnaby Road, Portsmouth, PO1 3FX, UK\\
$^{2}$Department of Physics and Astronomy, University of Waterloo, 200 University Ave W, Waterloo, ON N2L 3G1, Canada\\
$^{3}$Perimeter Institute for Theoretical Physics, 31 Caroline St. North, Waterloo, ON N2L 2Y5, Canada
}

\date{Accepted XXX. Received YYY; in original form ZZZ}

\pubyear{2018}

\begin{document}
\label{firstpage}
\pagerange{\pageref{firstpage}--\pageref{lastpage}}
\maketitle

\begin{abstract}
Redshift space distortions within voids provide a unique method to test for environmental dependence of the growth rate of structures in low density regions, where effects of modified gravity theories might be important. We derive a linear theory model for the redshift space void-galaxy correlation that is valid at all pair separations, including deep within the void, and use this to obtain expressions for the monopole $\xi^s_0$ and quadrupole $\xi^s_2$ contributions. Our derivation highlights terms that have previously been neglected but are important within the void interior. As a result our model differs from previous works and predicts new physical effects, including a change in the sign of the quadrupole term within the void radius. We show how the model can be generalised to include a velocity dispersion. We compare our model predictions to measurements of the correlation function using mock void and galaxy catalogues modelled after the BOSS CMASS galaxy sample using the Big MultiDark $N$-body simulation, and show that the linear model with dispersion provides an excellent fit to the data at all scales, $0\leq s\leq120\;h^{-1}$Mpc. While the RSD model matches simulations, the linear bias approximation does not hold within voids, and care is needed in fitting for the growth rate. We show that fits to the redshift space correlation recover the growth rate $f(z=0.52)$ to a precision of $2.7\%$ using the simulation volume of $(2.5\;h^{-1}\mathrm{Gpc})^3$.
\end{abstract}


\section{Introduction}
\label{sec:Introduction}

Galaxy redshift surveys provide a map of the large-scale structure of the Universe containing anisotropic distortions of the clustering caused by gravitationally-induced peculiar velocities that contribute to the galaxy redshifts, as first predicted by \citet{Kaiser:1987}. Measurement of these redshift space distortions (RSD) \citep[e.g.][]{Peacock:2001,Guzzo:2008,Beutler:2012,Samushia:2012,Reid:2012,Howlett:2015,Beutler:2017} can be used to determine the growth rate of structure, which can provide strong tests of General Relativity (GR). The theory of RSD in the galaxy clustering is however complicated by significant non-linear contributions which are important even at quite large pair separation scales, requiring sophisticated modelling \citep[e.g.][]{Scoccimarro:2004,Matsubara:2008,Taruya:2010,Reid:2011,Jennings:2011a}. 

The RSD modelling could in principle be simplified in underdense regions where the dynamics is closer to linear, such as in cosmic voids. Voids are large underdensities in the galaxy distribution which trace stationary points of the gravitational potential \citep{Nadathur:2017a}, where velocities are dominated by coherent bulk flows. Voids have been much studied recently in other contexts, including their action as weak gravitational lenses \citep{Krause:2013,Melchior:2014,Clampitt:2015,Sanchez:2016}, the secondary CMB anisotropies they generate through the integrated Sachs-Wolfe effect \citep[e.g.][]{Granett:2008a,Hotchkiss:2015a,Nadathur:2016b,Cai:2016b,Kovacs:2016} and the thermal Sunyaev-Zeldovich effect \citep{Alonso:2018}, and the Baryon Acoustic Oscillation (BAO) peak in their clustering \citep{Kitaura:2016b}. Velocity dynamics around voids have been studied using RSD in the void-galaxy cross-correlation \citep{Paz:2013,Hamaus:2015,Cai:2016a,Hamaus:2017a,Achitouv:2017a} and that of the void positions themselves \citep{Chuang:2017}.

In addition to possible simplifications in the linear modelling, an important advantage of studying RSD around voids is that it presents an opportunity to study the growth of density perturbations specifically in low-density regions. Several popular alternatives to GR, such as chameleon $f(R)$ gravity \citep{Hu:2007}, Dvali-Gabadadze-Porrati (DGP) \citep{Dvali:2000}, and Galileon (\citealt{Nicolis:2009}; \citealt{Deffayet:2009a}) gravity models, among others, make use of screening mechanisms in order to suppress fifth force effects in high density regions such as the Solar System; theories with such screening effects therefore predict environment-dependent differences in the growth rate $f$, which could be probed by the RSD effects within voids.

In this paper we study the void-galaxy cross-correlation function $\xi_{vg}$ in redshift space. \citet{Cai:2016a} have previously studied the same problem and provided a linear model for $\xi_{vg}$, which was subsequently also used by \citet{Hamaus:2017a}. However, this model correctly described simulation results only for void-galaxy pair separations greater than the void size, $r>R_v$, i.e. outside the low-density region of greatest interest. By extending the model using a form of the quasi-linear streaming model, \citet{Cai:2016a} were able to extend the region of validity slightly to $r>0.5R_v$, though it was still not correct in the void centres. In fact in the void centre region the model can lead to unphysical predictions of $\xi_{vg}<-1$. 

We revisit the derivation of the void-galaxy RSD model from first principles and identify linear-order terms that were neglected in the expression obtained by \citet{Cai:2016a} but are important in the void centre regions. These terms do not have counterparts in the linear Kaiser model for RSD in the galaxy correlation, and arise because of the restriction to void regions. They have important physical effects, in particular causing a change in sign of the quadrupole term within void regions. These terms become unimportant outside the void radius, where our expression matches that of \citet{Cai:2016a}. 

We show how the model can be extended to include a velocity dispersion without changing the linear nature of the theory. This is \emph{not} the same as the form of the Gaussian streaming model used in several other studies of the void-galaxy correlation \citep[e.g][]{Paz:2013,Hamaus:2015,Achitouv:2017a,Achitouv:2017b}. We show why it is not correct to use the standard streaming model result for the galaxy correlation in the void-galaxy case.

We then compare our theoretical results to data from mock void and galaxy catalogues from a large $N$-body simulation and show that our linear dispersion model provides an excellent fit to the data at all scales and performs significantly better than alternatives. We highlight the fact that the approximation of a linear galaxy bias does not hold within voids. We discuss the consequences of these results for obtaining an unbiased estimate of the growth rate within void environments. 

The layout of the paper is as follows. Section \ref{sec:simulation} describes the simulation data we use and the creation of galaxy and void mocks. In Section \ref{sec:theory} we derive the linear theory model for the void-galaxy correlation and its multipoles, and discuss differences with previous works. In Section \ref{sec:modelvdata} we compare theoretical predictions to simulation data and in Section \ref{sec:growthrate} we discuss strategies for measurement of the growth rate based on these results. We sum up and draw conclusions in Section ~\ref{sec:conclusions}.

\section{Data}
\label{sec:simulation}

\subsection{Simulation and galaxy mocks}
\label{sec:sims}

We use data from the $z=0.52$ redshift snapshot from the Big MultiDark (BigMD) $N$-body simulation \citep{Klypin:2016} from the MultiDark simulation project \citep{Prada:2012}. This simulation follows the evolution of $3840^3$ particles in a box of side $L=2500\;h^{-1}$Mpc using the {\small GADGET-2} \citep{Springel:2005} and Adaptive Refinement Tree \citep{Kravtsov:1997,Gottloeber:2008} codes, with cosmological parameters $\Omega_M=0.307$, $\Omega_B=0.048$, $\Omega_\Lambda=0.693$, $n_\mathrm{s}=0.95$, $\sigma_8=0.825$ and $h=69.3$. Initial conditions for the simulation were set using the Zeldovich approximation at starting redshift $z_i=100$. 

A halo catalogue was created for the given snapshot using the Bound Density Maximum algorithm \citep{Klypin:1997,Riebe:2013}. We populated these halos with mock galaxies using the Halo Occupation Distribution (HOD) model of \citet{Zheng:2007}, assigning central and satellite galaxies to halos according to a distribution based on the halo mass. Details of the algorithm and HOD model parameters used are described more fully in \citet{Nadathur:2017a}, who used the same mock catalogue: these parameters were taken from \citet{Manera:2013} and are designed to approximately reproduce the clustering and mean number density for galaxies in the Baryon Oscillation Spectroscopic Survey (BOSS) CMASS galaxy sample.

To measure dark matter (DM) densities in the simulation, we used a cloud-in-cell interpolation scheme to determine the DM density on a $2350^3$ grid using the full particle output of the simulation. We used this grid density field to determine the dark matter power spectrum $P(k)$. We then measured the galaxy power spectrum $P_{gg}(k)$ for the mocks; by fitting for the ratio $P_{gg}(k)/P(k)=b^2$ at large scales, $k\lesssim0.05\;h\mathrm{Mpc}^{-1}$, we determine the linear bias value for the galaxy mocks, $b=1.88$.

Using the real space galaxy positions $\mathbf{x}$ and velocities $\mathbf{v}$, we determine their redshift space positions in the plane-parallel approximation assuming the line of sight direction to be along the $z$-axis of the simulation box, 
\begin{equation}
\label{eq:s}
\mathbf{s} =\mathbf{x} + \frac{\mathbf{v}\cdot\hat{\mathbf{z}}}{aH}\,,
\end{equation}
where $a$ is the scale factor and $H$ is the Hubble rate.

\subsection{Void finding}
\label{sec:void-finding}

We identify voids in the real space galaxy mocks using the \zobov watershed void-finding algorithm \citep{Neyrinck:2008}. The \zobov algorithm uses a Voronoi tessellation field estimator (VTFE) technique to reconstruct the galaxy density field from the discrete distribution, and then identifies local minima in this field and the watershed basins around them, to form a non-overlapping set of voids corresponding to local density depressions. As in previous works \citep{Nadathur:2016a,Nadathur:2017a}, we define each individual density basin as a distinct void, without any additional merging of neighbouring regions. A fuller description of the algorithm and void properties can be found in \citet{Nadathur:2016a} and \citet{Nadathur:2017a}.

Although voids are of arbitrary shape and are in general far from spherically symmetric, it is convenient to define an effective spherical radius, $R_v=\left(3V/4\pi\right)^{1/3}$, where $V$ is the total volume of the void. We determine the centre of each void to be the centre of the largest sphere completely empty of galaxies that can be inscribed within the void \citep{Nadathur:2015b,Nadathur:2017a}. In Appendix \ref{appendix:centres} we consider the effect of defining the void centre as the volume-weighted barycentre of void member galaxies, as is also popular in the literature. Such a redefinition does not alter any of the qualitative conclusions in the following sections, but it decreases the available signal-to-noise for the RSD measurement and worsens the agreement with linear dynamics due to bulk velocity flows. 

It is important to stress that we apply the void-finding algorithm to the real space galaxy mocks and not to the shifted version in redshift space. As we discuss in the next section, this is crucial because all of the theoretical models for RSD in the void-galaxy cross-correlation discussed in this paper and elsewhere in the literature are based on assumptions that do not hold unless the real space void positions are known. \citet{Nadathur:2018b} show how this practical difficulty can be overcome when using survey data where the real space galaxy positions are not available.

Approximately 33000 voids are identified in the simulation box. As the void-finding algorithm is space-filling, these voids cover the entire box volume, and undoubtedly include some spurious identifications that do not correspond to genuine matter underdensities. We also do not expect a linear model of coherent velocity outflow from a void to successfully describe the RSD around very small voids, where the local environment of structures outside the void is important in determining the velocity field. Indeed \citet{Cai:2016a} find that RSD models for the void-galaxy correlation do not work for small voids. However, the distinction between `large' and `small' voids is somewhat ambiguous, and the numerical value of the cut  on void size depends both on the bias and number density of the galaxies in question \citep{Nadathur:2015b, Nadathur:2015c} as well as on the particular features of the voidfinder. 

We therefore restrict our void sample to the half of all voids with effective radius greater than the median radius. This is an easily reproducible criterion. For the mocks and voids used in this work, this means selecting $R_v\geq43\;h^{-1}$Mpc, which leaves $16\,421$ voids. Throughout the rest of this paper, we exclusively use this sample of voids. The mean effective void radius for this sample is $\overline{R_v}=55.6\;h^{-1}$Mpc. 

In some figures in later sections we show distances from the void centre in units of this mean radius, for context. However, we do not rescale distances in units of individual void sizes. Rescaling by individual void radii could improve the signal if the same features in the cross-correlation appear at the same rescaled distances for all voids, i.e. if the void size uniquely determines the void profile. The results of \citet{Nadathur:2015c,Nadathur:2015b,Nadathur:2017a} show that in general this is not the case, so we do not expect significant benefit from rescaling. On the other hand, rescaling effectively weights void-galaxy pair counts differently depending on the size of the void in the pair. This strongly up-weights the contribution of small voids relative to large ones in the average cross-correlation, which is undesirable, as the model is expected to work better for larger voids. It also significantly complicates the error determination.

Note that restricting our sample to the largest $50\%$ of voids need not necessarily be the optimal choice to ensure validity of linear theory. \citet{Nadathur:2017a} show that environmental effects around voids are more strongly correlated with a combination of void size and density than with void size alone, so this may provide a better selection criterion. However, the median size cut implemented here has the advantage of simplicity, and as we show later, is sufficient that a purely linear RSD model already provides an excellent fit to the data.

\subsection{Measuring correlation functions}
\label{sec:correlation}

Given the discrete void and galaxy populations in our simulation, measurement of the void-galaxy correlation is simply a matter of counting void-galaxy pairs as a function of the separation between them, which we do using a version of the {\small CUTE} correlation function code \citep{Alonso:CUTE}.\footnote{\url{http://members.ift.uam-csic.es/dmonge/CUTE.html}} That is, we measure the cross-correlation 
\begin{equation}
\label{eq:xi}
\xi = \frac{D_v D_g}{N_v \overline{n}_g V} -1,
\end{equation}
where $D_vD_g$ is the total number of void-galaxy pairs within a given separation bin of volume $V$, $N_v$ is the total number of voids, and $\overline{n}_g$ is the mean number density of galaxies, $N_g/V_\mathrm{box}$. Note that as we do not rescale distances based on individual void radii, for a given separation bin the volume $V$ is the same for all voids, which means we weight each void-galaxy pair equally. As the distribution of voids and galaxies is uniform throughout the simulation box, Eq. \ref{eq:xi} is equivalent to the use of the Landy-Szalay estimator \citep{Landy:1993} in the limit of using infinitely many random points.

All correlation functions $\xi$ are measured in the same way. By using purely radial bins in the separation $r$ we obtain the monopoles $\xi(r)$, and by binning also with respect to angle we obtain $\xi(r,\mu)$, where $\mu$ is the cosine of the angle to the line of sight direction. For all correlation function measurements we use 50 equally-spaced radial bins of width $\Delta r=2.4\;h^{-1}$Mpc out to a maximum separation of $120\;h^{-1}$Mpc. For angular measurements we use 100 angular bins in the range $0\leq\mu\leq1$. Quadrupoles of the correlation function are obtained from $\xi(r,\mu)$ by 
\begin{equation}
\label{eq:quad_defn}
\xi_2(r) = 5\,\int_0^1 \xi(r,\mu)P_2(\mu)d\mu\,,
\end{equation}
where $P_2(\mu)=\frac{1}{2}(3\mu^2-1)$ is the Legendre polynomial of order 2.

We also require the stacked DM density profile $\delta(r)$ around voids, which is equivalent to the void-matter cross-correlation. As the direct computation of pair counts for voids and DM particles is too computationally expensive given the size of the simulation, we instead interpolate the DM particle output on to a $2350^3$ density grid using a CIC interpolation scheme, and use the grid values to determine the stacked average void density profile. 

\section{Theory}
\label{sec:theory}

\subsection{The void-galaxy cross-correlation in redshift space}
\label{sec:base model}

Let $\mathbf{X}$ denote the comoving location of a void centre, and $\mathbf{x}$ the location of a galaxy in its vicinity. The real-space separation vector for the void-galaxy pair, $\mathbf{r}=\mathbf{x}-\mathbf{X}$, is transformed to $\mathbf{s}$ in redshift space. \emph{Assuming that the total number of void-galaxy pairs is unchanged  by the shift to redshift space}, we require that
\begin{equation}
\label{eq:void-galaxy conservation}
\left(1+\xi^s_{vg}(\mathbf{s})\right) d^3s = \left(1+\xi^r_{vg}(\mathbf{r})\right) d^3r\,,
\end{equation}
where $\xi_{vg}$ denotes the void-galaxy cross-correlation (in what follows we will suppress the subscript where there is no risk of confusion), and the supersripts $^s$ and $^r$ denote redshift and real space respectively. $\xi_{vg}$ may also be viewed as the galaxy number density profile around the void, and is sometimes denoted $\delta_g$ in this context.

In the distant-observer approximation, the void-galaxy separation vector transforms in redshift space as
\begin{equation}
\label{eq:full coordinates}
\mathbf{s} = \mathbf{r} + \frac{\left(\mathbf{v}-\mathbf{V}\right)\cdot\hat{\mathbf{X}}}{aH}\hat{\mathbf{X}}\,,
\end{equation}
where $\mathbf{v}$ denotes the peculiar velocity of the galaxy and $\mathbf{V}$ the effective velocity of the void centre. The appearance of $\mathbf{V}$ in this equation reflects the fact that in general the void centre positions will also shift under the RSD mapping \citep{Chuang:2017}.

However, modelling $\mathbf{V}$ is complicated, for reasons discussed below. In common with previous works, we therefore further assume that \emph{the void centre position is invariant under the RSD mapping}. This allows the simplification 
\begin{equation}
\label{eq:coordinates}
\mathbf{s} = \mathbf{r} + \frac{\mathbf{v}\cdot\hat{\mathbf{X}}}{aH}\hat{\mathbf{X}}\,,
\end{equation}
which now depends only on the galaxy peculiar velocity rather than the pairwise velocity.

Note that both the assumptions made here are known to fail if voids are identified directly using the redshift space galaxy field! \citet{Zhao:2016}, \citet{Chuang:2017} and \cite{Nadathur:2018b} show that void numbers are not conserved in redshift space. \citet{Chuang:2017} showed that the autocorrelation of the positions of redshift-space voids themselves shows an RSD pattern, which means that the mapping from real to redshift space does not preserve void positions. Thus for such voids the coordinate transformation $\mathbf{r}\rightarrow\mathbf{s}$ cannot depend only on the galaxy peculiar velocities $\mathbf{v}$ as assumed in Eq. \ref{eq:coordinates}. As a result, any model for the void-galaxy RSD which is derived from the assumption inherent in Eqs. \ref{eq:void-galaxy conservation} and \ref{eq:coordinates} cannot be consistently applied to redshift-space voids. 

As pointed out by \citet{Chuang:2017}, at small to intermediate scales void-finding constitutes a non-linear transformation of the density field, and therefore necessarily leads to a void velocity bias different from unity \citep{Seljak:2012}. To model $\mathbf{V}$ it would be necessary to determine this velocity bias accurately. However, the action of the void-finder algorithm cannot be described by a simple mathematical model. This means the non-linear transformation and the void velocity bias would need to be determined empirically.

\citet{Nadathur:2018b} show that the violation of these two assumptions for redshift-space voids leads to strong additional contributions to the observed RSD pattern well within the scale of the mean void radius, meaning that it affects even the 1-void term in the cross-correlation \citep{Cai:2016a}. Accounting for these effects in the modelling by modifying Eqs. \ref{eq:void-galaxy conservation} and \ref{eq:coordinates} appears challenging. On the other hand, if only real-space void positions are used for the cross-correlation, both assumptions are satisfied by construction: void numbers are necessarily conserved and their positions are by definition unchanged by the RSD mapping. Therefore, for real space voids both Eqs. \ref{eq:void-galaxy conservation} and \ref{eq:coordinates} \emph{are} valid.

In this paper we are interested in developing the correct theoretical model for this more manageable case. We therefore apply the following algorithm:
\begin{enumerate}
    \item we identify voids using the real space galaxy field,
    \item we cross-correlate the real space void and galaxy positions to determine $\xi^r(r)$,
    \item we cross-correlate the real space void positions with the redshift space galaxies to determine $\xi^s(\mathbf{s})$.
\end{enumerate}
This procedure is simple to apply for the simulation we consider, since real space galaxy positions are known. For survey data this will not be the case. However, \citet{Nadathur:2018b} show how in this case the real-space void positions can be effectively recovered using a Zeldovich reconstruction technique, allowing fair comparison of the data with the model derived here. The reconstruction procedure will be applicable for any galaxy surveys designed for BAO detection.

We therefore proceed with the derivation from Eqs. \ref{eq:void-galaxy conservation} and \ref{eq:coordinates}. From Eq.~\ref{eq:coordinates}, the line-of-sight component of the separation vector is
\begin{equation}
\label{eq:coordinates LOS}
\mathbf{s}_{||} = \mathbf{r}_{||} + \frac{\mathbf{v}_{||}}{aH}.
\end{equation}
The restriction to real-space voids means that from symmetry arguments, the average coherent velocity flow must be spherically symmetric and directed along the radial direction, $\mathbf{v}=v_r(r)\hat{\mathbf{r}}$. (Note that this is also \emph{not} the case for redshift-space voids, \citealt{Nadathur:2018b}.) The determinant of the Jacobian for the coordinate transformation is therefore
\begin{equation}
\label{eq:Jacobian basic}
\left| J\left(\frac{\mathbf{s}}{\mathbf{r}}\right)\right| = 1+\frac{v_r}{raH} + \frac{\left(v_r^\prime-v_r/r\right)}{aH}\mu^2\,,
\end{equation}
where $^\prime$ denotes the derivative with respect to the radial distance $r$, and $\mu$ is the cosine of the  angle between the line-of-sight direction and the separation vector,
\begin{equation}
\label{eq:mu defn}
\mu \equiv \frac{\mathbf{X}\cdot\mathbf{r}}{|\mathbf{X}||\mathbf{r}|}=\cos\theta.
\end{equation}
Eq.~\ref{eq:void-galaxy conservation} can therefore be rewritten as
\begin{equation}
\label{eq:xis basic}
1+\xi^s(\mathbf{s}) = \left(1+\xi^r(\mathbf{r})\right)\left[ 1+\frac{v_r}{raH} + \frac{\left(v_r^\prime-v_r/r\right)}{aH}\mu^2 \right]^{-1}.
\end{equation}

%==================Fig.: =======================%
\begin{figure*}
\begin{center}
\includegraphics[scale=0.52]{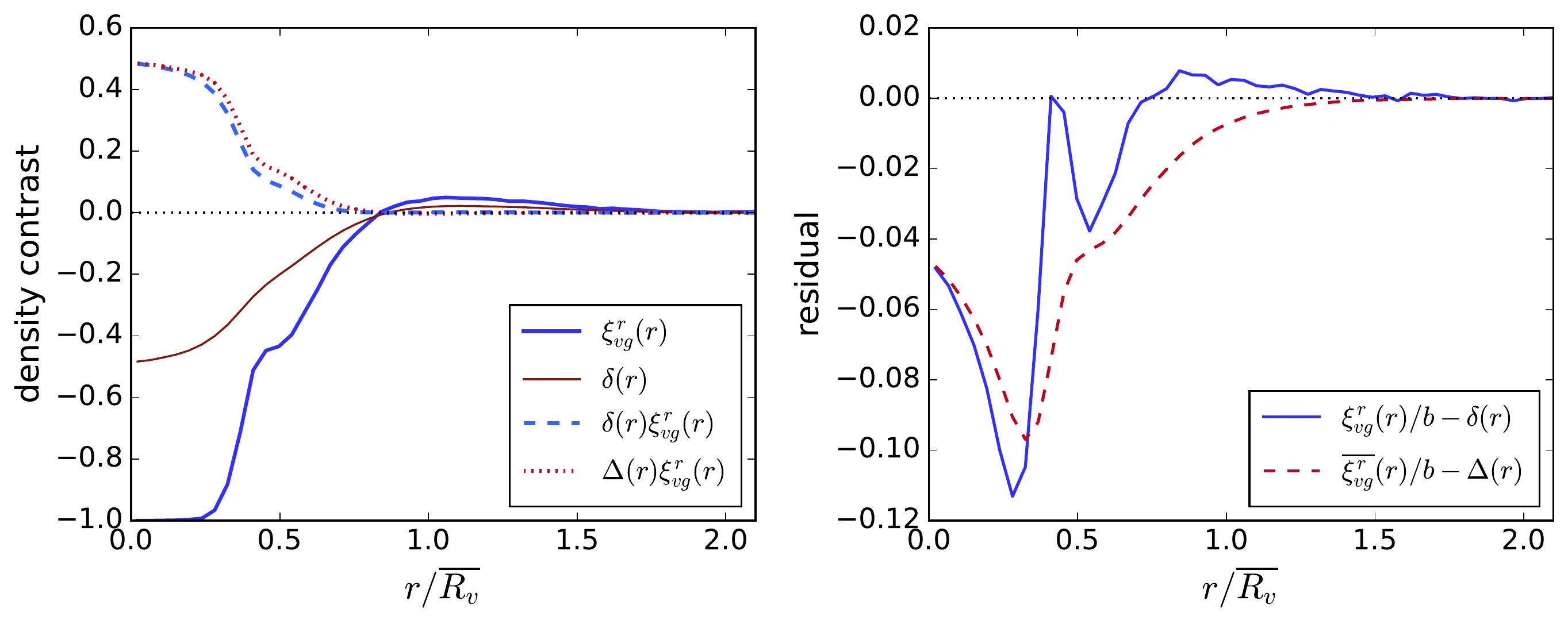}
\caption{\emph{Left}: The stacked average real space void-galaxy cross-correlation $\xi^r_{vg}$ (thick solid, blue) and void density contrast $\delta$ (thin solid, red) as functions of the distance from the centre of the void, for voids in our simulation sample. The dashed and dot-dashed lines show the products $\xi\delta$ and $\xi\Delta$. Note the kink in $\xi^r(r)$ around $r\sim0.5\overline{R_v}$. A similar though less pronounced kink is present in $\delta(r)$ at the same location. \emph{Right}: The residual between the linear bias approximated value $\xi^r_{vg}(r)/b$, where bias $b$ is determined from the large-scale clustering (see text), and the true mass density contrast $\delta(r)$ in the simulation (blue solid line). The red dashed curve shows the corresponding residual $\overline{\xi^r_{vg}}(r)/b-\Delta(r)$ for the integrated quantities.}
\label{fig:approximations}
\end{center}
\end{figure*}
%==================Fig.:=======================%

We will assume that the peculiar velocity field is coupled to the density as in linear dynamics, and is determined by the void alone, so that 
\begin{equation}
\label{eq:linear velocity}
\mathbf{v}(\mathbf{r}) = -\frac{1}{3}faH\Delta(r)\mathbf{r}\,,
\end{equation}
where $\Delta(r)$ is the average mass density contrast within radius $r$ of the void centre,
\begin{equation}
\label{eq:Delta defn}
\Delta(r)\equiv\frac{3}{r^3}\int_0^r \delta(y)y^2 dy\,,
\end{equation}
where $\delta(r)$ is the mass density profile of the void, and $f=\mathrm{d}\ln D/\mathrm{d}\ln a$, with $D$ the growth factor and $a$ the scale factor, is the linear growth rate of density perturbations.

Note that the assumption of linear dynamics in Eq. \ref{eq:linear velocity} is not crucial to the development of the model. \citet{Achitouv:2017b} assumed a specific non-linear coupling between density and velocity, obtaining a different expression for $v_r(r)$. So long as the radial velocity field is spherically symmetric, as is the case for real-space voids, Eq. \ref{eq:xis basic} must hold for any form of the velocity profile $v_r(r)$. In Section \ref{sec:modelvdata} we examine the validity of the assumption in Eq. \ref{eq:linear velocity}.
 
Combining Eqs.~\ref{eq:xis basic} and \ref{eq:linear velocity}, and using the fact that $\Delta^\prime(r)=3/r\left(\delta(r)-\Delta(r)\right)$, we obtain
\begin{equation}
\label{eq:xis invJ}
1+\xi^s(\mathbf{s}) = \left(1+\xi^r(r)\right)\left[ 1 - \frac{f}{3}\Delta(r) - f\mu^2\left(\delta(r)-\Delta(r)\right)\right]^{-1}\,.
\end{equation}
The expansion of the second term on the RHS must be performed with care. For the voids in highly biased galaxy tracers that we consider in this paper, the mass density contrast is small enough that we can safely drop terms of second order or higher in $\delta$ and $\Delta$. Note that this is separate from the assumption of the validity of linear dynamics, Eq.~\ref{eq:linear velocity}, and may not hold for other void populations, even if the dynamics is close to linear. 

However, independent of assumptions about the size of $\delta$ and $\Delta$, terms of order $\xi^r\delta$ and $\xi^r\Delta$ \emph{cannot} be assumed to be small. The very fact that voids are identified by selecting regions with very few or no galaxies \emph{ensures} that in the void interior $\xi\sim-1$, and thus that terms proportional to $\xi\delta$ and $\xi\Delta$ are comparable to $\delta$ and $\Delta$ and thus must be retained to linear order. The left panel of Figure~\ref{fig:approximations} demonstrates this explicitly for the voids in our simulation. In fact for void centre definition used here, $\xi^r(r)$ is exactly $-1$ for small $r$, since the centres of voids contain no galaxies. Thus $\xi^r\delta=-\delta$ and it is obvious these terms are not negligible. This conclusion is however independent of the void centre choice. In Appendix \ref{appendix:centres} we consider a different definition of the void centre, using which $\xi^r(r)\geq-0.8$ at all $r$, and show that the terms $\xi\delta$ and $\xi\Delta$ are still comparable in size to $\delta$, $\Delta$.

Retaining these terms in the expansion, Eq.~\ref{eq:xis invJ} therefore reduces to 
\begin{multline}
\label{eq:xis full}
\xi^s\left(s,\mu\right) = \xi^r(r)+\frac{f}{3}\Delta(r)\left(1+\xi^r(r)\right) \\ 
+f\mu^2\left[\delta(r)-\Delta(r)\right]\left(1+\xi^r(r)\right) \,,
\end{multline}
where the difference between real space and redshift space coordinates $r$ and $s$ is
\begin{equation}
\label{eq:svr full}
r = s\left(1+\frac{f}{3}\Delta(s)\mu^2\right)\,,
\end{equation}
to the same linear order in $\Delta$. The effect of this coordinate shift is neglected in the linear Kaiser theory for RSD in the galaxy correlation \citep{Kaiser:1987} because it only appears at second order in that derivation \citep[see, e.g., Section 4.2 of][]{Hamilton:1998}. However, in our case, the coordinate shift appears in $\xi$ at linear order, via the Taylor expansion 
\begin{equation}
\label{eq:Taylor}
\xi(r) = \xi(s) + \xi^\prime(s)\frac{f}{3}s\Delta(s)\mu^2 + \ldots\,.
\end{equation}

The effect of the terms in the model above can be understood as follows. The velocity outflow from voids introduces a mapping between the real space correlation $\xi^r$ at separation $r$ and the redshift space correlation $\xi^s$ at separation $s$, where $r<s$ along the line of sight, for $\Delta<0$. Since $\xi$ decreases towards the centre, this results in a `stretching' of the void along the line of sight direction, in accordance with naive intuition for the RSD effect around underdensities. This effect is most important where the gradient of $\xi^r$ is steepest.

On the other hand, as noted and explained by \citet{Cai:2016a}, the terms $f\Delta/3+f\mu^2\left(\delta-\Delta\right)$, arising from the Jacobian of the transformation, give rise instead to a `squashing' along the line of sight.\footnote{In principle it is possible for these terms to also lead to a stretching for pathological choices of $\delta(r)$, but in such a case it is unlikely that a linear model based on Eq.~\ref{eq:linear velocity} would be viable anyway. In practice, we do not observe such behaviour for any subset of the voids in our simulation.} This effect is suppressed by a factor of $(1+\xi^r)$ and is therefore most important around the void edges and outside the void. The competing effects of the stretching and squashing terms can be seen in Figure~\ref{fig:comparison}. The combination results in a change in the sign of the quadrupole term within the void, going from negative to positive with increasing $s$, as will be shown in Section~\ref{sec:modelvdata}.

\subsection{Linear bias approximation}
\label{sec:bias}

Eqs.~\ref{eq:xis full} and \ref{eq:svr full} together provide the correct expression for the void-galaxy cross-correlation in redshift space assuming linear dynamics and at linear order in $\delta$, $\Delta$. They form the base model that we will use to explain the simulation data. However, their use requires knowledge of the true mass density contrast $\delta(r)$ and its integral version $\Delta(r)$, which cannot be directly determined for voids in survey data. An approximation that is often made \citep{Cai:2016a,Hamaus:2017a} is to assume that a simple linear bias relationship holds within the voids, so that $\xi^r(r) = b\delta(r)$ and $\overline{\xi^r}(r) \equiv 3/r^3\int_0^r \xi^r(y)y^2 dy = b\Delta(r)$, where $b$ is the large-scale linear bias factor determined from the galaxy clustering.

Under this approximation, the full model of Eqs.~\ref{eq:xis full} and \ref{eq:svr full} can be rewritten as 
\begin{multline}
\label{eq:xis bias}
\xi^s\left(s,\mu\right) = \xi^r(r)+\frac{\beta}{3}\overline{\xi^r}(r)\left(1+\xi^r(r)\right)  \\
+\beta\mu^2\left[\xi^r(r)-\overline{\xi^r}(r)\right]\left(1+\xi^r(r)\right)\,,
\end{multline}
where
\begin{equation}
\label{eq:svr bias}
r = s\left(1+\frac{\beta}{3}\overline{\xi^r}(s)\mu^2\right)\,,
\end{equation}
and $\beta\equiv f/b$.

To explore the validity of this linear bias approximation within voids, we determine the linear bias factor for our mock galaxies by computing the (real space) galaxy power spectrum $P_{gg}(k)$ and the full matter power spectrum $P(k)$ for the simulation box, and  fitting $P_{gg}(k) = b^2P(k)$ at large scales, $k\lesssim0.05\,h^{-1}$Mpc. For this value of $b$, the residuals $\xi^r(r)/b-\delta(r)$ and $\overline{\xi^r}(r)/b - \Delta(r)$ compared to the linear bias approximation are shown in the right panel of Figure~\ref{fig:approximations}. Importantly, the deviations are (i) scale dependent, and (ii) large. For the void centre definition we use, the linear bias assumption leads to fractional differences of $\sim25\%$ from the true value within the void radius. In Appendix \ref{appendix:centres}, we show that residuals remain large and scale-dependent irrespective of the void centre choice.

It is important to emphasise that the deviation from the linear bias relationship seen can arise purely due to a \emph{statistical selection effect} caused by the void-finding process, independent of non-linear or environment-dependent bias models (which we do not consider here). Where a linear bias $b$ characterises the galaxy clustering as above, the relationship between the local galaxy overdensity $\delta_g$ and the matter overdensity $\delta$ is still $\delta_g = b\delta +\epsilon$, where $\epsilon$ represents a stochastic term. If the stochastic term is unbiased, the conditional expectation value over the entire universe or simulation box is indeed
\begin{equation}
\label{eq:meanbias}
\langle\delta_g|\delta\rangle = b\delta\,.
\end{equation}
However, voids are by construction selected as relatively rare regions with a small galaxy density. The cross-correlation within void regions is thus equivalent to a constrained 1-point function. This necessarily introduces a selection effect in $\langle\delta|\delta_g\equiv\xi^r_{vg}\rangle$: regions with negative fluctuations in $\epsilon$ are more likely to be selected as voids than those with positive $\epsilon$, thus biasing the mean relationship such that $\xi^r/b$ systematically overestimates the depth of the void where $\xi^r<0$. Conversely,  $\xi^r/b$ overestimates the height of the wall around the void edge where $\xi^r>0$. This will be true \emph{even when $b=1$}, i.e. when the tracers used to identify the voids and determine $\xi$ are a random subset of the DM particles in the simulation\footnote{Under certain simplifying assumptions about the nature of the stochastic term, this selection effect can be modelled analytically; see, e.g., \cite{Gruen:2016troughlensing}, Sec. 3.1. In general, the void selection condition will introduce correlations between the stochastic terms at different radial distances, complicating the modelling. We will not pursue this further in the current work.} as explicitly demonstrated in \citet{Nadathur:2015b}. The selection effect will however be much reduced if the tracer profile $\xi^r$ is measured using a different set of tracers to those used to select void locations, such as an overlapping galaxy sample with higher mean number density, or the set of all halos in the simulation from which the mock galaxy hosts are drawn.

For our purposes, the consequences of this failure of the linear bias relationship within voids are two-fold:
\begin{enumerate}
\item if the same galaxy tracers are used to identify the voids and to measure the RSD pattern, Eqs.~\ref{eq:xis bias} and \ref{eq:svr bias} are \emph{necessarily} always inaccurate with respect to the true model in Eqs.~\ref{eq:xis full} and \ref{eq:svr full} in the void interiors, but
\item the extent of the discrepancy is somewhat mitigated by the fact that the discrepancy from the linear bias value is largest in the regions where the $(1+\xi^r(r))$ term in Eq.~\ref{eq:xis bias} is small. 
\end{enumerate}

For the mocks we have both the matter and galaxy density fields, so we can contrast results for the real to redshift space mapping using the directly measured density field within voids as well as that inferred from the linear bias assumption, as done in Section \ref{sec:modelvdata}. This allows us to isolate the effect of this assumption on the RSD model.

\subsection{Multipole expansion}
\label{sec:multipoles}

It is convenient to expand $\xi^s$ in terms of its multipoles,
\begin{equation}
\label{eq:multipoles}
\xi^s_\ell(s) = \int_0^1 \xi^s(s,\mu)\left(1+2\ell\right)P_\ell(\mu)d\mu\,
\end{equation}
where $P_\ell(\mu)$ are the Legendre polynomials of order $\ell$. At linear order in $\delta$ and $\Delta$, only the monopole and quadrupole terms are non-zero. Using $P_0(\mu)=1$ and $P_2(\mu)=(3\mu^2-1)/2$, these can be calculated by direct integration for the model using either Eqs.~\ref{eq:xis full} and \ref{eq:svr full} or Eqs.~\ref{eq:xis bias} and \ref{eq:svr bias}. For all numerical calculation of model multipoles presented in this paper, we use this approach.

However, approximate analytical forms can also be obtained by using the first two terms of the expansion in Eq.~\ref{eq:Taylor}, to rewrite Eq.~\ref{eq:xis full} as
\begin{multline}
\label{eq:xis approx}
\xi^s\left(s,\mu\right) \simeq \xi^r(s)+\frac{f}{3}\Delta(s)\left(1+\xi^r(s)\right) \\ 
+f\mu^2\left[\delta(s)-\Delta(s)\right]\left(1+\xi^r(s)\right) +\frac{f\mu^2}{3}s\xi^{r\prime}(s)\Delta(s)\,,
\end{multline}
to linear order in $\delta$, $\Delta$. Substituting this into Eq.~\ref{eq:multipoles} gives monopole
\begin{equation}
\label{eq:monopole}
\xi^s_0(s) = \xi^r(s) + \frac{f}{9}s\,\xi^{r\prime}(s)\Delta(s) + \frac{f}{3}\delta(s)\left[1+\xi^r(s)\right]\,,
\end{equation}
and quadrupole
\begin{equation}
\label{eq:quadrupole}
\xi^s_2(s) = \frac{2f}{9}s\,\xi^{r\prime}(s)\Delta(s) +\frac{2f}{3}\left[\delta(s)-\Delta(s)\right]\left[1+\xi^r(s)\right]\,.
\end{equation}
Corresponding versions of these equations can be obtained when also including the linear bias approximation.

A consequence that follows from these expressions is that the quadrupole-to-monopole ratio does not factorize conveniently as in the model of \citet{Cai:2016a} and so cannot be used as an estimator for the growth rate.

\subsection{Velocity dispersion and the streaming model}
\label{sec:dispersion}
 
So far we have used a pure Kaiser model to describe $\xi^s$: that is, we assumed that velocities around void centres exactly follow the coherent outflow described by the linear relationship in Eq.~\ref{eq:linear velocity}. In Section~\ref{sec:modelvdata} we will show that this is a surprisingly good approximation for the mean outflow velocity on all scales, so the model of Eq.~\ref{eq:xis full} provides a good qualitative description of the void-galaxy correlation seen in simulation.

However, to enable a quantitative fit to the data, a more realistic model must account for the dispersion of galaxy velocities around this mean. To do this, we introduce a dispersion in the line-of-sight galaxy velocities, such that
\begin{equation}
\label{eq:velocity disp}
\mathbf{v} = v_r\hat{\mathbf{r}} + v_{||}\hat{\mathbf{X}}\,,
\end{equation}
where $v_r$ is the coherent radial component given by Eq.~\ref{eq:linear velocity}, and $v_{||}$ is a zero-mean random variable with probability distribution function $P(v_{||})$. This results in an integral for $\xi^s$:
\begin{eqnarray}
\label{eq:xis dispersion}
1+\xi^{s,d}(\sigma,\pi) &=& \int dv_{||} P(v_{||})\left(1+\xi^s\left(\sigma,\pi-v_{||}/aH\right)\right),  \nonumber \\
&=& \int dv_{||} P(v_{||})(1+\xi^r\left(r\right))\left| J\left(\frac{\mathbf{s}}{\mathbf{r}}\right)\right|^{-1},
\end{eqnarray}
where $\sigma$ and $\pi$ are respectively distances transverse to and along the line of sight, $r=\sqrt{r_\sigma^2+r_\pi^2}$, with $r_\sigma=\sigma$ and $r_\pi = \pi - (v_{||}+v_r\mu)/aH$, and $\left|J\left(\frac{\mathbf{s}}{\mathbf{r}}\right)\right|$ is as in Eq.~\ref{eq:Jacobian basic}. The effect of the dispersion in $v_{||}$ is primarily through shuffling the radial and transverse distances contributing to $\xi^r$ in the integral. The dispersion has a negligible effect on the Jacobian, as the contribution to the radial outflow averages to zero.

We will take the probability distribution function $P(v)$ to have a Gaussian form,
\begin{equation}
\label{eq:Pv}
P(v) = \frac{1}{\sqrt{2\pi}\sigma_v}\exp\left(-\frac{v^2}{2\sigma_v^2}\right).
\end{equation}
The dispersion is most generally a function of the radial separation scale, $\sigma_v=\sigma_v(r)$. The validity of the Gaussian assumption for $P(v)$ may also vary with $r$. We consider these questions in more detail in Section~\ref{sec:modelvdata}. 

Note that Eq.~\ref{eq:xis dispersion} is not quite the same as an adaptation of the streaming model \citep[e.g.,][]{Fisher:1995,Scoccimarro:2004,Reid:2011} to the void-galaxy case. In particular, as the coherence of density and velocity fields is accounted for by the inverse Jacobian, this is still fundamentally a \emph{linear} model in all senses. We still assume the coherent outflow $v_r$ is determined by linear dynamics as before; in evaluating the Jacobian we still drop terms higher than linear order in $\delta$ and $\Delta$; and the convolution with $P(v)$ merely broadens the coordinate shift effect already present at linear order in Eqs.~\ref{eq:xis full} and \ref{eq:svr full}. In the limit of zero dispersion, where $P(v_{||})\rightarrow\delta_D(v_{||})$, this model reduces to Eq.~\ref{eq:xis full}. Eq.~\ref{eq:xis dispersion} is thus similar to the dispersion model used for the galaxy correlation \citep{Hamilton:1998} before the development of the full streaming model, except that the width of the velocity distribution is allowed to depend on scale. 

Unlike in the case for the galaxy correlation, this linear dispersion model is already sufficient to provide an excellent fit to the data at all scales, as we show in Section \ref{sec:xis}. We may therefore safely leave the development of a full streaming model for the void-galaxy case to future work.

\subsection{Comparison with previous results in the literature}
\label{sec:comparison}

The basic linear model we have presented above differs in several key respects to other models for the void-galaxy correlation in the literature \citep[e.g.][]{Paz:2013,Cai:2016a,Hamaus:2017a,Achitouv:2017a,Achitouv:2017b}. In Section \ref{sec:modelvdata} we will compare these models to $\xi^s$ measured in the simulation data and show that our model provides a significantly better description of the true void-galaxy correlation. Before doing that, however, it is instructive to examine the reasons for the difference in the derivations.

\citet{Cai:2016a} \citep[and subsequently][]{Hamaus:2017a} follow a derivation similar to that described in Section~\ref{sec:base model}, with three important differences: (1) they do not include terms of order $\xi\delta$ and $\xi\Delta$ in the expansion of Eq.~\ref{eq:xis invJ}, (2) they approximate $s=r$ contrary to Eq.~\ref{eq:svr full}, and (3) they assume the linear bias approximation (Section~\ref{sec:bias}) holds within voids.\footnote{\citet{Cai:2016a} actually provide expressions assuming $\xi^r=\delta$ and $\overline{\xi^r}=\Delta$, which is equivalent to using the linear bias assumption with $b=1$.}  The first two assumptions alone give rise to 
\begin{equation}
%\label{eq:xis Cai}
\xi^s(s,\mu) = \xi^r(s) + \frac{f}{3}\Delta(s) + f\mu^2\left[\delta(s)-\Delta(s)\right]\,,
\end{equation}
which can be compared to Eq. \ref{eq:xis full}, and the addition of the linear bias assumption gives
\begin{equation}
\label{eq:xis Cai}
\xi^s(s,\mu) = \xi^r(s) + \frac{\beta}{3}\overline{\xi^r}(s) + \beta\mu^2\left[\xi^r(s)-\overline{\xi^r}(s)\right]\,,
\end{equation}
which can be compared to Eq. \ref{eq:xis bias}. 

It is worth emphasising that this model is derived from \emph{exactly the same} assumptions as those described in Section \ref{sec:base model}, i.e., void conservation (Eq. \ref{eq:void-galaxy conservation}), invariance of void positions (Eq. \ref{eq:coordinates}), the spherical symmetry of the velocity outflow, and linear dynamics for the velocity field (Eq. \ref{eq:linear velocity}). The only differences arise from the truncation of the series expansion of the square brackets in Eq. \ref{eq:xis basic} to exclude terms of order $\xi\delta$ and $\xi\Delta$. In this sense, the model of \citet{Cai:2016a} is an approximation to the more complete model in Eq. \ref{eq:xis full}. If the terms of order $\xi\delta$ and $\xi\Delta$ were truly small, both models would lead to the same predictions. However, since they are not, neglecting these terms leads to large differences in the final model predictions, as we show. 
 
The model for $\xi^s$ in Eq. \ref{eq:xis Cai} can be decomposed into a monopole term
\begin{equation}
\label{eq:monopole Cai}
\xi^s_0(s) = \left(1+\frac{\beta}{3}\right)\xi^r(s)\,,
\end{equation}
and a quadrupole
\begin{equation}
\label{eq:quadrupole Cai}
\xi^s_2(s) = \frac{2\beta}{3}\left[\xi^r(s)-\overline{\xi^r}(s)\right]\,.
\end{equation}

A simple heuristic way to see that these expressions cannot be correct within the void interior is to consider the limiting case where $\xi^r\sim-1$ close to the void centre, at $s\sim0$. Here Eq.~\ref{eq:monopole Cai} predicts a redshift space monopole $\xi^s_0<-1$, which is unphysical. This is a simple consequence of the premature truncation of the series expansion, dropping terms that are not small. More generally, Eq.~\ref{eq:xis Cai} is a poor description of $\xi^s(\mathbf{s})$ everywhere that $(1+\xi^r)$ is small, which in practice for the voids considered here means at least for all radial separations within the mean void radius. In Appendix \ref{appendix:centres} we show that this statement is independent of the choice of void centres.

In addition, because of the assumption that $s=r$, Eq.~\ref{eq:xis Cai} qualitatively misses the important stretching effect along the line of sight due to the mapping of void-galaxy separations from real to redshift space. As we show in the next section, this causes a change in the sign of the quadrupole term within the void radius which is not captured by Eq. \ref{eq:quadrupole Cai}. In fact this change of sign was already seen in simulation data by \citet{Cai:2016a} (see Figures 1-3 in their paper), although not satisfactorily explained. This shortcoming of the model is the primary reason why the growth rate estimator proposed in that paper does not work for $s<\overline{R_v}$, leading to negative reconstructed values of the growth rate within voids \citep[see the discussion in][]{Cai:2016a}.

On the other hand, in contrast to both our results and those of \citet{Cai:2016a}, \citet{Hamaus:2017a} do not observe any change of sign in the quadrupole $\xi^s_2$ measured in BOSS galaxy data and associated galaxy mocks. This is because they use voids identified using the galaxy positions in redshift space -- when this is done, neither of the fundamental assumptions used to obtain Eqs.~\ref{eq:void-galaxy conservation} and \ref{eq:coordinates} are valid, and so neither our model nor that of \citet{Cai:2016a} -- nor any of the other theoretical models for $\xi^s$ elsewhere in the literature -- will satisfactorily describe the data. In particular, the application of such models to the cross-correlation with redshift-space voids leads to a strongly biased reconstruction of the fiducial growth rate (see the demonstration in Appendix A of \citealt{Nadathur:2018b}). 

In a separate work, we discuss a solution to this problem via the use of a Zeldovich reconstruction technique to identify real-space void positions \citep{Nadathur:2018b}.

Finally, a number of different authors \citep[e.g.][]{Paz:2013,Hamaus:2015,Hamaus:2016,Cai:2016a,Achitouv:2017a,Achitouv:2017b} have used an analogy with the Gaussian streaming model for the galaxy autocorrelation to model $\xi^s$:
\begin{equation}
\label{eq:GSM}
1+\xi^s(\sigma,\pi) = \int \frac{\left(1+\xi^r(r)\right)}{\sqrt{2\pi}\sigma_v}\exp\left(-\frac{(v_{||}-v_r(r)\mu)^2}{2\sigma_v^2}\right)\,dv_{||},
\end{equation}
with $r=\sqrt{\sigma^2+\left(\pi-v_{||}/aH\right)^2}$. Comparison with Eq.~\ref{eq:xis dispersion} shows that this expression differs from our dispersion model. Unlike our dispersion model, it does not reduce to the linear expression, Eq.~\ref{eq:xis full}, in the limit of zero dispersion when $\mathbf{v}=v_r\hat{\mathbf{r}}$.\footnote{This model does approximately match Eqs.~\ref{eq:xis full} and \ref{eq:xis dispersion} when the conditions $|\xi^r|,\,|\delta|\,,|\Delta|\ll1$ are satisfied and the dispersion is small, $\sigma_v\ll raH$ \citep{Cai:2016a}. In practice this holds only outside the void boundary, $s\gtrsim1.5\overline{R_v}$.} Also unlike our dispersion model, it provides a poor fit to the data, as we will show in Section \ref{sec:modelvdata}.

The reason for this is simple: the streaming model in Eq.~\ref{eq:GSM} has been derived for the case of linear, Gaussian fluctuations in the galaxy or matter \emph{autocorrelation} \citep{Fisher:1995,Scoccimarro:2004}, and cannot simply be extended by analogy to the void-galaxy cross-correlation. Not only is the moment generating function $\mathcal{Z}(\lambda,\mathbf{r})$ \citep[see][]{Scoccimarro:2004} different for the void-galaxy case, the derivation of the Gaussian streaming model explicitly assumes that the only non-zero cumulant of the density field is the second. However, the void-galaxy case involves constrained averages of the fields at specially selected locations rather than over all space, so that by definition the expectation value $\langle\delta\rangle\neq0$ within the void. As a result the analogy is invalid and Eq.~\ref{eq:GSM} does not describe the correct streaming model for the void-galaxy cross-correlation. 

\section{Simulation results}
\label{sec:modelvdata}

\subsection{Void density profiles}
\label{sec:density}

The angle-averaged stacked void density profiles $\delta(r)$ and $\xi^r(r)$ are equivalent to the monopoles of the real-space void-DM and void-galaxy cross-correlations, which are measured as described in Section \ref{sec:correlation}.  To determine $\Delta(r)$ and $\overline{\xi^r}(r)$ we interpolate $\delta(r)$ and $\xi^r(r)$ respectively and numerically evaluate the corresponding integrals.

The profiles obtained for our void population are shown in the let panel of Figure~\ref{fig:approximations}. Here for context we show radial separations in units of the mean radius of all voids in the stack, $\overline{R_v}=55.6\;h^{-1}$Mpc. As discussed in Section~\ref{sec:theory}, in the interior region of voids the terms $\xi\delta$ and $\xi\Delta$ are of comparable magnitude to $\delta$ and $\Delta$, meaning that they must be included in the expansion of Eq. \ref{eq:xis basic}. The right panel of Figure~\ref{fig:approximations} shows the residuals between the true void matter density profiles and those inferred from applying the linear bias approximation to $\xi^r(r$), as discussed in Section \ref{sec:bias}.

\subsection{Velocity profiles and dispersion}
\label{sec:velocity}

%==================Fig.: =======================%
\begin{figure}
\begin{center}
\includegraphics[scale=0.42]{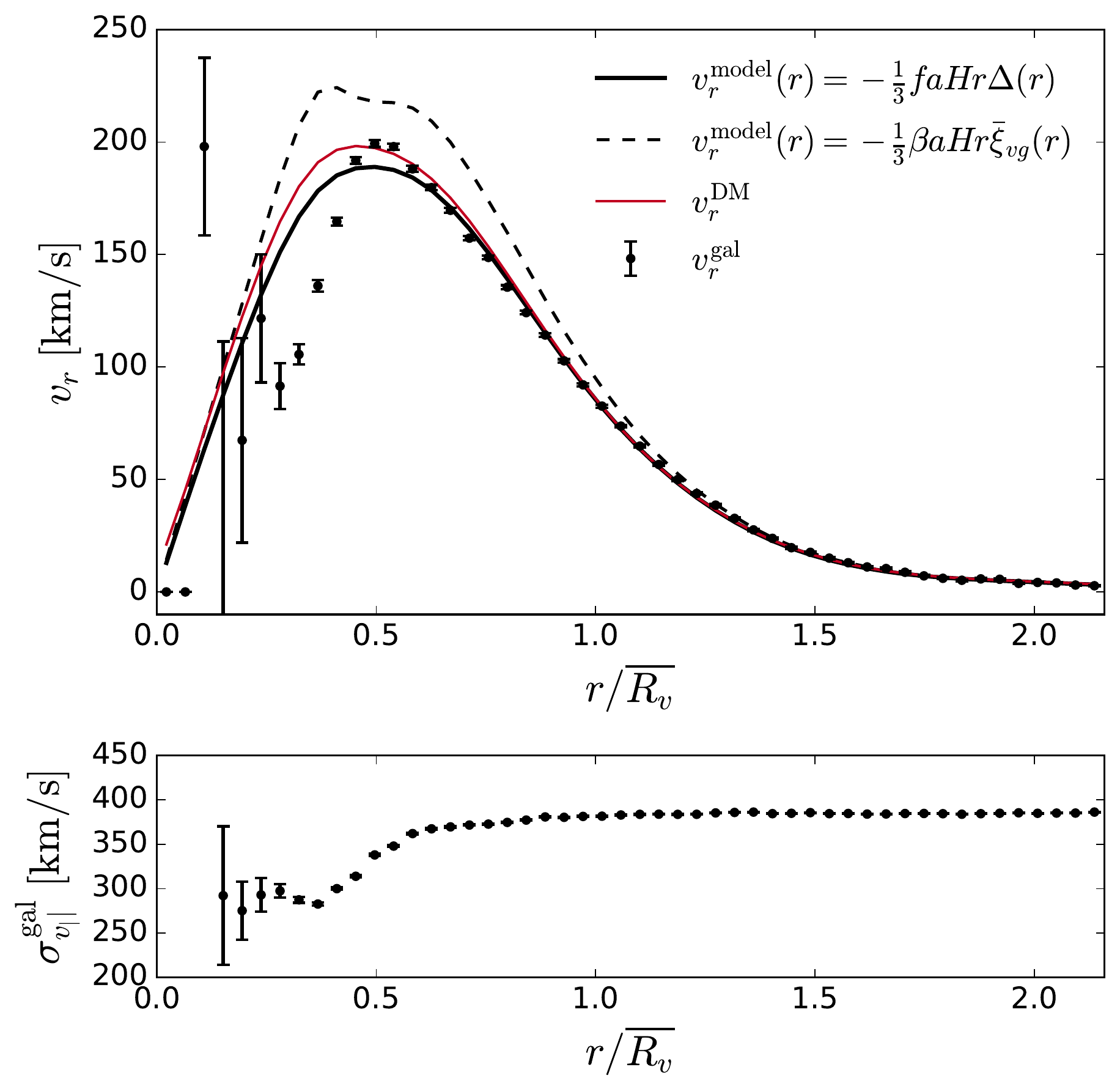}
\caption{\emph{Top}: Stacked average galaxy radial velocity profiles around void centres for voids in our simulation (data points with error bars), compared to the correct linear theory prediction (thick solid line), and the linear theory prediction if the constant linear galaxy bias assumption is also used (dashed line). The red (thin, solid) curve shows the measured average radial velocity profile of dark matter halos around the same voids. The error bars on this curve are much smaller than those for galaxies, so are omitted for clarity. \emph{Bottom}: The measured dispersion of galaxy line-of-sight velocities $\sigma_{v_{||}}^\mathrm{gal}(r)$.}
\label{fig:velocity}
\end{center}
\end{figure}
%==================Fig.:=======================%

%==================Fig.: =======================%
\begin{figure*}
\begin{center}
\includegraphics[scale=0.6]{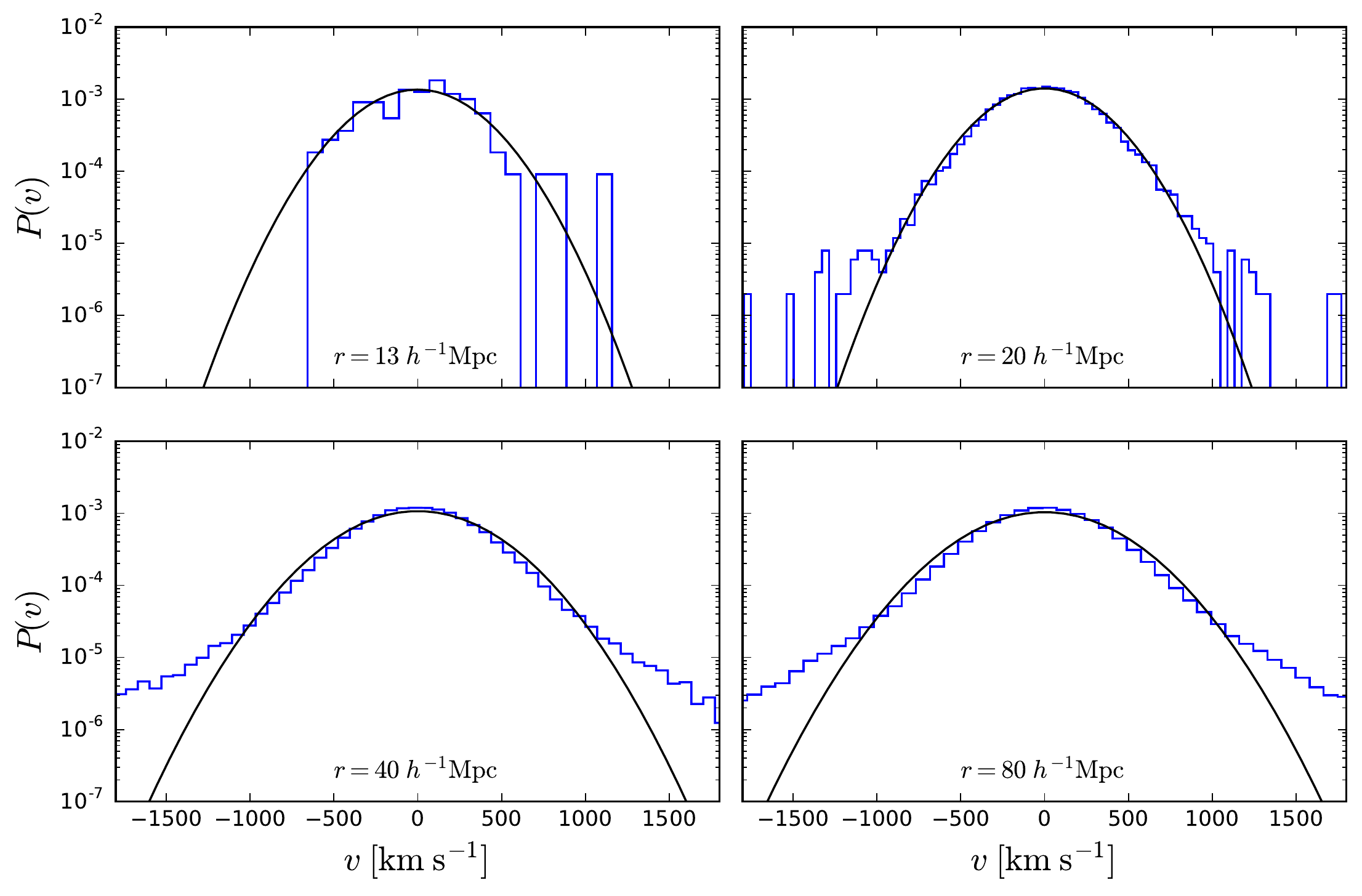}
\caption{Normalised histograms (blue) for the line-of-sight velocity component for galaxies in four bins of the radial void-galaxy separation distance, as indicated in the text insets. The black solid curves represent Gaussian distributions with widths equal to the corresponding values of $\sigma_{v_{||}}^\mathrm{gal}(r)$ in Figure~\ref{fig:velocity}.}
\label{fig:dispersion}
\end{center}
\end{figure*}
%==================Fig.:=======================%

We measure the average (stacked) galaxy velocity profile around the void centres in the simulation as
\begin{equation}
\label{eq:avgvel}
v_r^\mathrm{gal}(r) = \frac{1}{N_{vg}}\sum_{i,j}\mathbf{v}_j\cdot\hat{\mathbf{r}}_{ij}\,,
\end{equation}
where $\mathbf{r}_{ij}$ is the real space separation vector between the $i$th void and $j$th galaxy and $\mathbf{v}_{j}$ the velocity of that galaxy, and the sum runs over all $N_{vg}$ void-galaxy pairs with $|\mathbf{r}_{ij}|$ in the range $(r-dr,r+dr)$. As before, we use $50$ equally-spaced bins out to a maximum separation of $120\;h^{-1}$Mpc. We also measure the dispersion in the line-of-sight velocity component, $\sigma_{v_{||}}^\mathrm{gal}$, defined by
\begin{equation}
\label{eq:sigmav}
\sigma_{v_{||}}^\mathrm{gal}(r) = \left[\frac{1}{N_{vg}}\sum_{i,j}\left(\mathbf{v}_j\cdot\hat{\mathbf{X}}_j - \overline{v_{||}}(r)\right)^2\right]^{1/2}\,,
\end{equation}
where $\hat{\mathbf{X}}_j$ is the line of sight direction, which is the same for all galaxies in the plane-parallel approximation and taken to be along the $z$-axis of the simulation box, $\overline{v_{||}}(r)$ is the mean line of sight velocity component at $r$, and the sum is over all void-galaxy pairs in the given separation bin as before. 

Note that Eqs.~\ref{eq:avgvel} and \ref{eq:sigmav} weight each void-galaxy pair equally. An alternative way to define these quantities, which has sometimes been used in the literature, is to calculate the average and dispersion separately for each void, and then average the results over all voids. This is equivalent to weighting each void equally: this procedure leads to smaller measured values of $v_r^\mathrm{gal}$ and $\sigma_{v_{||}}^\mathrm{gal}$ at small separations, since most voids do not have any galaxies in interior bins and thus contribute zero to the average. At large $r$ both methods will largely agree. However, for comparison with the void-galaxy cross-correlation, which also weights each void-galaxy pair equally, our method is the more appropriate.

Finally, to control for possible statistical effects due to the void selection criterion as seen in the relationship between $\xi^r(r)$ and $\delta(r)$, we also measure the average DM velocity profile around void centres. To do this we use the same procedure as for measuring $\delta(r)$: the Cartesian components $v_x$, $v_y$ and $v_z$ of the DM particle velocities are interpolated onto $2350^3$ grids using a CIC interpolation scheme, and the resulting gridded velocity fields are used to evaluate the stacked average radial DM velocity profiles around void centres, $v_{r}^\mathrm{DM}(r)$.   

Figure~\ref{fig:velocity} plots the results as a function of distance from the void centre, together with the linear velocity model for $v_r(r)$ obtained from Eq.~\ref{eq:linear velocity} (solid black line). The linear model is calculated using the fiducial value of the growth rate for the given simulation and redshift, $f=0.761$. The first conclusion that can be drawn from this is that linear dynamics gives an \emph{extremely good} description for the mean $v_{r}^\mathrm{DM}(r)$, with deviations at $\lesssim10\%$ over the entire range of scales tested \citep[see also][]{Hamaus:2014a}. It is also a good model for the mean galaxy radial velocity profile $v_{r}^\mathrm{gal}(r)$ at distances $r\gtrsim30\;h^{-1}$Mpc, where $v_{r}^\mathrm{DM}(r)$ and $v_{r}^\mathrm{gal}(r)$ coincide. 

Closer to the void centres, $r<\overline{R_v}$, $v_{r}^\mathrm{gal}(r)$ starts to deviate from both the linear model and $v_{r}^\mathrm{DM}(r)$. While some scatter is expected where the errors in $v_{r}^\mathrm{gal}(r)$ become large due to the small number of void-galaxy pairs (note that the two interior-most bins have no galaxies at all), in the range $10\lesssim r\lesssim30\;h^{-1}$Mpc the difference is statistically significant. This coincides with the `kink' visible in the $\xi^r(r)$ profile in Figure~\ref{fig:approximations}, which arises due to the fact that the void centre definition means that for each void, there are 4 galaxies at the same distance from the void centre. A similar, though less pronounced, kink is also present at the same distance in $\delta(r)$. This suggests that the velocity shortfall at this distance is a physical consequence of some aspect of the void selection algorithm. We leave a fuller explanation of this effect for future work.

Recently, \citet{Achitouv:2017b} proposed a semi-empirical non-linear correction term to modify the linear velocity relationship of Eq.~\ref{eq:linear velocity}, which effectively reduces the predicted outflow term $v_r(r)$ within voids. Figure~\ref{fig:velocity} shows that such a correction term is disfavoured in our data, since $v_{r}^\mathrm{DM}(r)$ is very close to the linear prediction and never less than it. However, note that incorrectly assuming a linear bias for the density profiles, $\Delta =\overline{\xi^r}/b$ in Eq.~\ref{eq:linear velocity} (shown by the dashed line in Figure~\ref{fig:velocity}) does indeed consistently overestimate the true velocity outflow. This can be understood by reference to Figure~\ref{fig:approximations}, which shows that assuming a linear bias overestimates the true matter deficit within the void. 

%==================Fig.: =======================%
\begin{figure*}
\begin{center}
\includegraphics[scale=0.55]{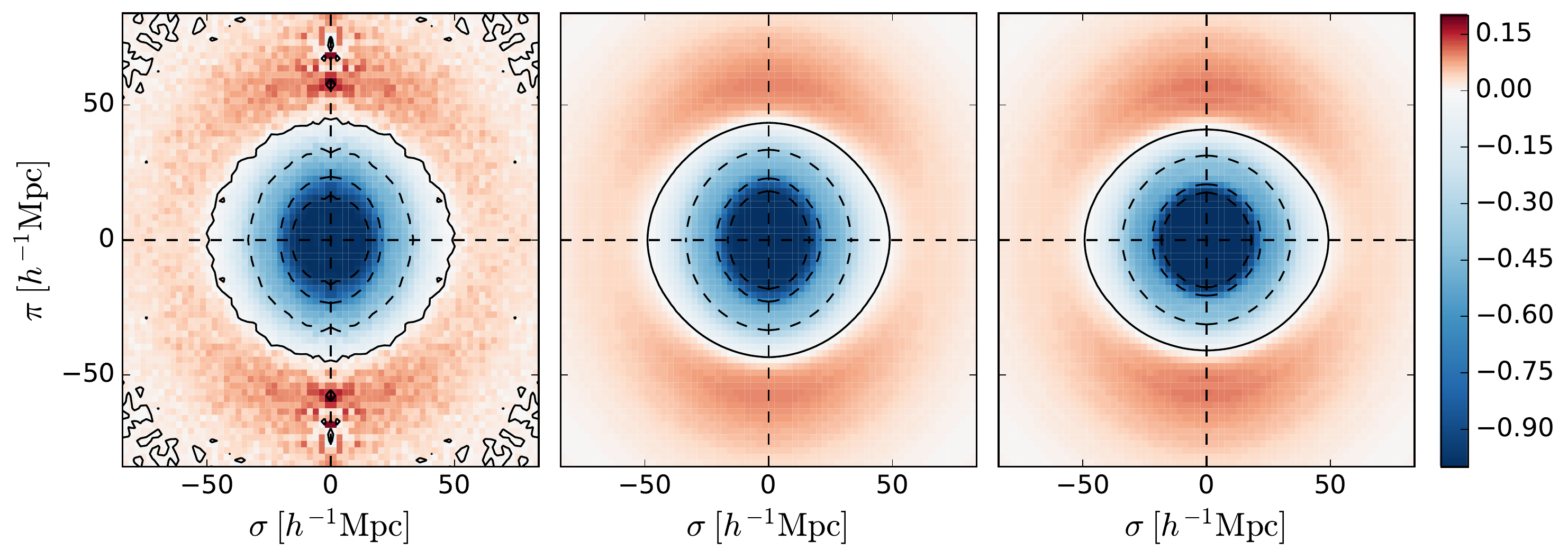}
\caption{\emph{Left}: The 2D void-galaxy cross-correlation function $\xi^s(\sigma,\pi)$ in redshift space measured for voids in the simulation. Dashed curves indicate contour lines for $\xi^s=-0.9$, $-0.6$ and $-0.3$, the solid curve is the contour $\xi^s=0$. \emph{Centre}: The corresponding theoretical prediction for our linear model. Eq. \ref{eq:xis full}. \emph{Right}: The model prediction for the model of \citet{Cai:2016a}, Eq. \ref{eq:xis Cai}. Note that in this case the colour scale is saturated at the centre, as the model predicts $\xi^s<-1$.}
\label{fig:comparison}
\end{center}
\end{figure*}
%==================Fig.:=======================%

%==================Fig.: =======================%
\begin{figure*}
\begin{center}
\includegraphics[scale=0.55]{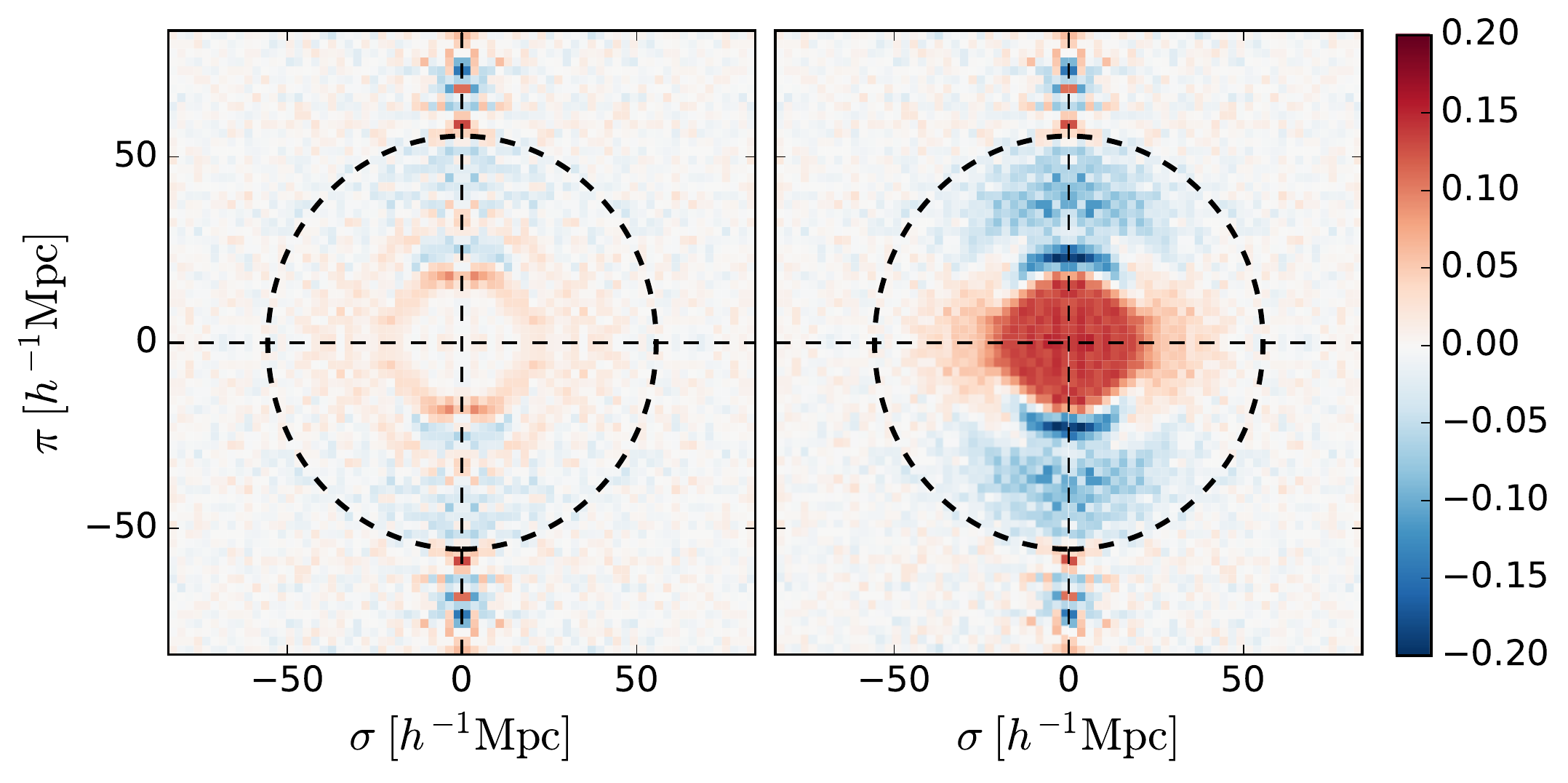}
\caption{Residual differences $\xi^s(\sigma,\pi)-\xi^s_\mathrm{th}(\sigma,\pi)$ between the measured void-galaxy cross-correlation and theoretical predictions. \emph{Left}: For the model in Eq.~\ref{eq:xis full}. \emph{Right}: For the model in Eq.~\ref{eq:xis Cai}. The thick dashed circle in both plots indicates the mean void radius $\overline{R_v}$.}
\label{fig:residuals}
\end{center}
\end{figure*}
%==================Fig.:=======================%

The lower panel of Figure~\ref{fig:velocity} shows the velocity dispersion $\sigma_{v_{||}}^\mathrm{gal}$ as a function of radial distance. Errors are estimated assuming velocities follow a normal distribution. The dispersion is approximately constant outside the void radius but decreases in the interior region, with a sharp drop in the same region as the measured radial velocity starts to deviate from the linear prediction. In the interiormost bins, the small number of void-galaxy pairs means the dispersion cannot be reliably estimated, so these points are excluded.

Where the dispersion $\sigma_{v_{||}}^\mathrm{gal}$ can be measured, it is always significantly larger than the mean radial velocity. This is in agreement with \citet{Achitouv:2017b} but differs from the results of \citet{Hamaus:2015} and \citet{Cai:2016a}, who report a much smaller dispersion within the mean void radius. It is possible this is due to the use of equal weighting for each void rather than each void-galaxy pair in the latter two papers.

Figure~\ref{fig:dispersion} shows the measured probability distribution function for line of sight velocities in four different bins of $r$, compared to the Gaussian distribution assumed in Eq.~\ref{eq:Pv} with $\sigma_v$ set to the corresponding value of $\sigma_{v_{||}}^\mathrm{gal}$. The distribution is symmetric around zero at all scales, and close to Gaussian at small scales, although at large separation scales exponential wings are seen which deviate from Gaussianity. At these large scales the impact of dispersion itself is small. Therefore the Gaussian assumption, while not strictly correct, still provides a good fit---as we show later. Note that this is quite different from the case for the distribution of pairwise velocities for galaxies, which is highly skewed at small scales and far from Gaussian at all scales \citep{Scoccimarro:2004}.

\subsection{Redshift space correlation function}
\label{sec:xis}

%==================Fig.: =======================%
\begin{figure}
\begin{center}
\includegraphics[scale=0.5]{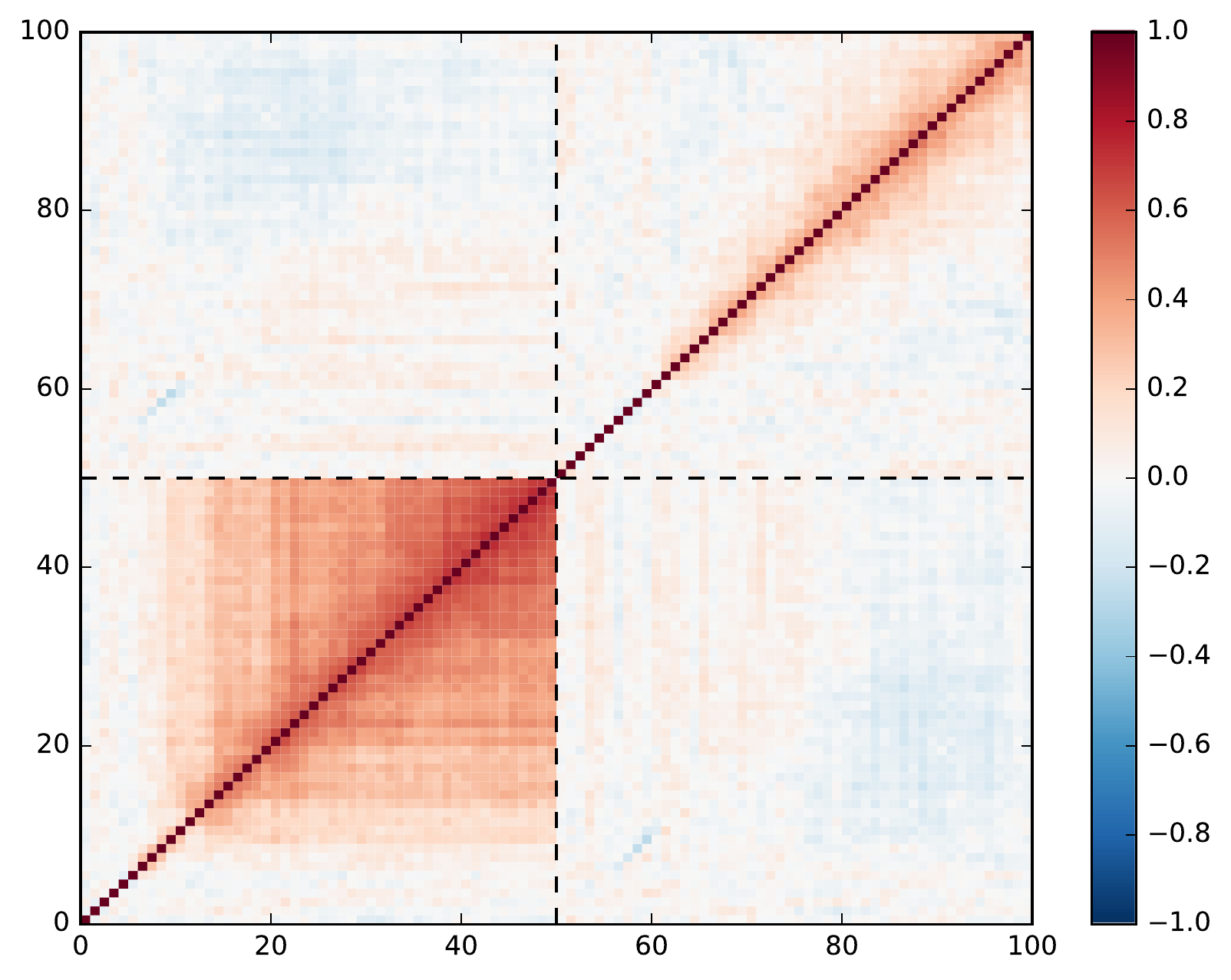}
\caption{Cross-correlation coefficients of the covariance matrix for multipoles of $\xi^s$, $\mathbf{C}_{ij}$, defined in Eq.~\ref{eq:covmat}. The dashed black lines divide the monopole (lower left block) and quadrupole (upper right) contributions.}
\label{fig:covmat}
\end{center}
\end{figure}
%==================Fig.:=======================%

%==================Fig.: =======================%
\begin{figure*}
\begin{center}
\includegraphics[scale=0.4]{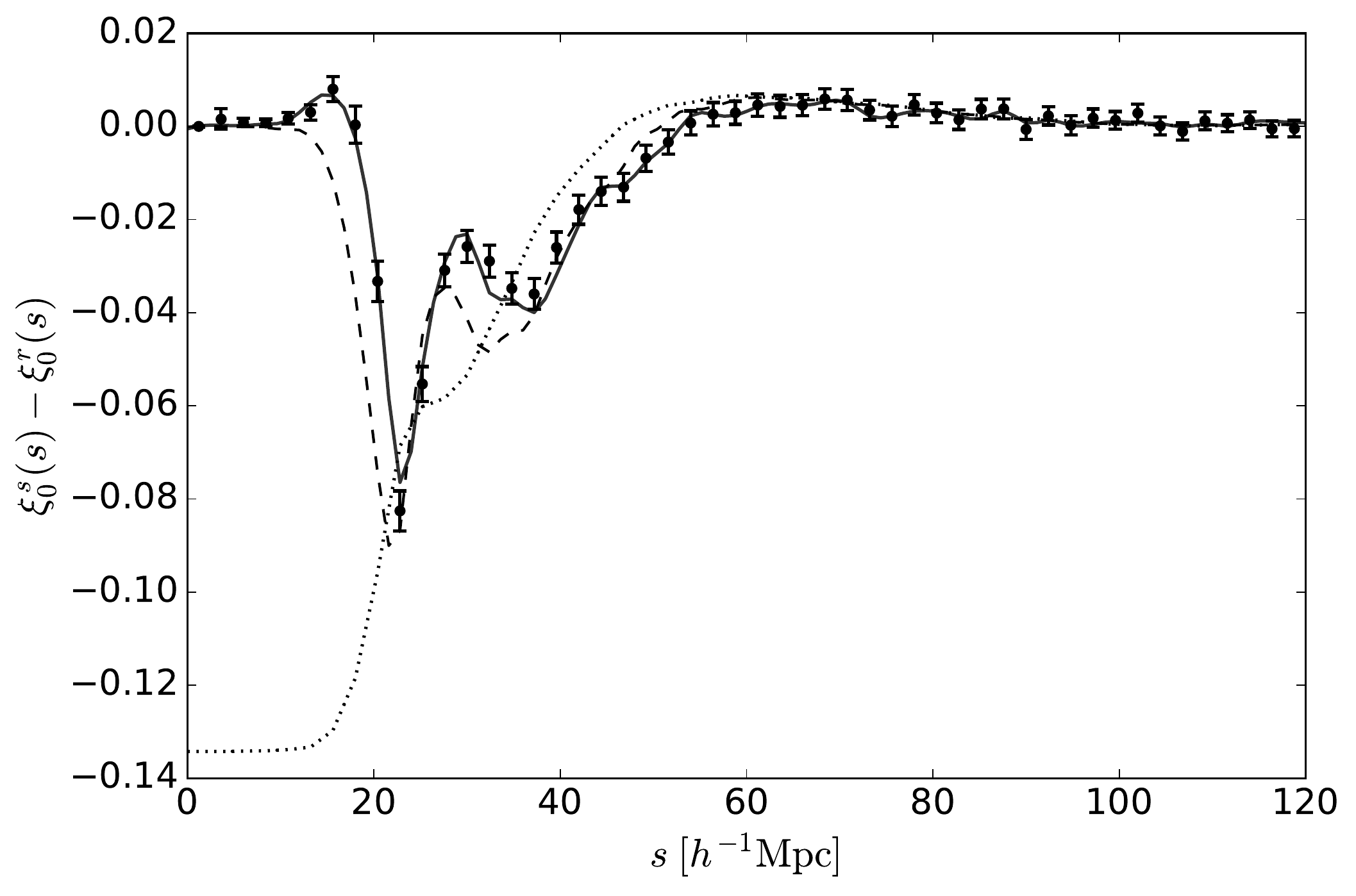}
\includegraphics[scale=0.4]{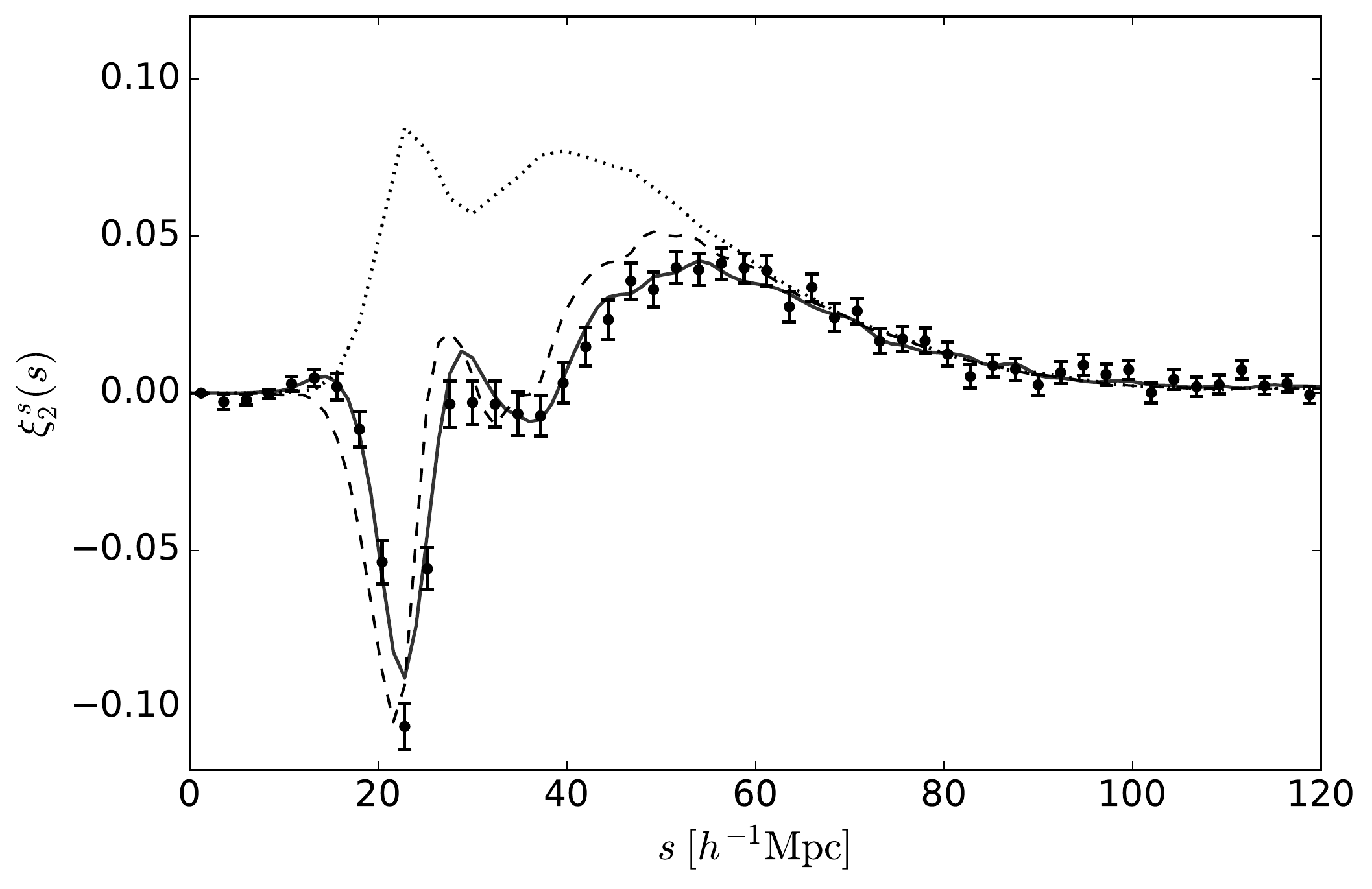}
\caption{Measured multipoles of the redshift space void-galaxy cross-correlation $\xi^s$ as a function of the radial void-galaxy separation $s$ (data points). Error bars are derived from diagonal elements of the estimated covariance matrix. \emph{Left}: The monopole $\xi^s_0(s)$. For visual clarity, this is plotted as the difference $\xi^s_0-\xi^r$ compared to the real space monopole, shown in Figure \ref{fig:approximations}. \emph{Right}: The quadrupole $\xi^s_2(s)$. In both panels, the solid curve shows the theoretical prediction for our full linear model including velocity dispersion, using Eq. \ref{eq:xis dispersion}; the dashed curve is the prediction for the linear model without the velocity dispersion term, using Eq. \ref{eq:xis full}; and the dotted curve is the prediction for the incomplete model of Eq. \ref{eq:xis Cai}.}
\label{fig:model multipoles}
\end{center}
\end{figure*}
%==================Fig.:=======================%

Figure~\ref{fig:comparison} shows the measured redshift space cross-correlation $\xi^s(\sigma,\pi)$ as a function of the transverse and line-of-sight distances, compared to the prediction from our model, Eq.~\ref{eq:xis full}, and that of the model of \citet{Cai:2016a}, Eq.~\ref{eq:xis Cai}. The qualitative features visible in the data can be understood on the basis of the equations derived in Section~\ref{sec:base model} above. Close to the void centre, the measured $\xi^s$ shows approximate spherical symmetry. At intermediate distances, the effect of the coordinate shift discussed in Section~\ref{sec:base model} causes an elongation effect, stretching the contours of equal $\xi^s$ along the line of sight direction. At large distances, $s\simeq\overline{R_v}=55\;h^{-1}$Mpc, this effect is reversed, with a relative squashing of the contours of $\xi^s(\sigma,\pi)$ along the line of sight direction. 

To better understand the relative merits of the two theoretical models, Figure~\ref{fig:residuals} shows the residual difference between the measured $\xi^s(\sigma,\pi)$ and the predicted correlation functions in the two cases. It is clear that the model of \citet{Cai:2016a} is a very poor fit to the data at all radial separations within the mean void radius $\overline{R_v}$, and particularly so very close to the centre, where Eq.~\ref{eq:xis Cai} predicts unphysical values of $\xi^s<-1$. By contrast, the inclusion of the $\left(1+\xi^r(r)\right)$ factors in Eq.~\ref{eq:xis full} correctly ensures that the theory prediction matches the measured value at the void centre---where in fact RSD effects are absent, $\xi^s(s)\simeq\xi^r(s)$---and also produces a much better fit to the data at all distances within the void interior.

Nevertheless, Figure~\ref{fig:residuals} shows that even for the improved model, when the velocity dispersion term is not included the theory residuals are still potentially significant. To perform a quantitative analysis, we instead perform a multipole expansion of the angular correlation function $\xi^s(s,\mu)$ from the simulation data to extract the monopole and quadrupole terms using Eq.~\ref{eq:multipoles} for comparison with theory. The multipoles are measured in 50 radial bins in the range $0\leq s\leq120\;h^{-1}$Mpc, and we use 100 bins in $\mu$. Note that we do \emph{not} rescale radial distances by the void radius, thus using the same bin sizes for each void. Such a rescaling would constitute a strong assumption of self-similarity in void profiles, which is not justified \citep{Nadathur:2015b,Nadathur:2015c,Nadathur:2017a}. 

To estimate the error in this measurement we use a jack-knife resampling technique. We divide our simulation box into $N_{s}=512$ non-overlapping cubic sub-boxes, each measuring $312.5\;h^{-1}$Mpc on a side, and determine the correlation function and multipoles excluding each sub-box in turn, which we combine into the data vector $\mathbf{y}^{(k)}=(\xi^{(k)}_0,\xi^{(k)}_2)$, for $k=1,\ldots,N_s$. The covariance matrix is then determined as
\begin{equation}
\label{eq:covmat}
\mathbf{C}_{ij} = \frac{N_s -1}{N_s}\sum_{k=1}^{N_s}\left(y^{(k)}_i-\overline{y_i}\right)\left(y^{(k)}_j-\overline{y_j}\right)\,.
\end{equation}
Figure~\ref{fig:covmat} shows the resulting matrix of cross-correlation coefficients. For the monopole, this matrix is far from diagonal, showing strong correlation between bins even at large separations. For the quadrupole it is closer to diagonal, although correlations between neighbouring bins are still significant, especially at large separations $s$. The $\chi^2$ values for the fit of a given theoretical model to the data are calculated as
\begin{equation}
\label{eq:chi2}
\chi^2 = \sum_{ij} \left(y^\mathrm{th}_i-y_i\right) \mathbf{C}_{ij}^{-1} \left(y^\mathrm{th}_j-y_j\right)\,.
\end{equation}

Figure~\ref{fig:model multipoles} shows the measured values of the monopole $\xi^s_0(s)$ and quadrupole $\xi^s_2(s)$ from the simulation data, compared to theory predictions derived in the previous section evaluated using the fiducial growth rate $f=0.761$. For visual clarity, the monopole values are shown as the difference with respect to the real space version, $\xi^s_0-\xi^r$. Two important features are immediately apparent: the quadrupole changes sign within the void interior, with a negative dip at intermediate values of $s$ changing to a broad $\xi^s_2(s)>0$ feature at $s\simeq\overline{R_v}=55\;h^{-1}$Mpc; and the redshift space monopole differs from the real space version only at intermediate $s$, with $\xi^s_0-\xi^r\simeq0$ both at the very centre (where $\xi^r\simeq-1$) and around the mean void radius. 

The dashed curves in each panel show the predictions of the basic linear model derived from Eq.~\ref{eq:xis full} without accounting for velocity dispersion: these capture the main qualitative features above but do not provide a good fit to the data, as expected from visual inspection of the residuals in Figure~\ref{fig:residuals}. The solid curves show the prediction for the linear dispersion model of Eqs.~\ref{eq:xis dispersion} and \ref{eq:Pv}, where we have used the radial dependence of the dispersion $\sigma_v(r)$ measured in the data, shown in Figure~\ref{fig:velocity}. This model provides an excellent fit to the data \emph{at all scales}: the $\chi^2/\mathrm{d.o.f.}$ values for this model are $41.1/48$ for the monopole alone, $54.3/48$ for the quadrupole alone, and $121.3/98$ for both combined. 

Finally, the dotted curves in Figure~\ref{fig:model multipoles} show the predictions for $\xi^s_0$ and $\xi^s_2$ according to the model presented by \citet{Cai:2016a,Hamaus:2017a}, Eqs.~\ref{eq:monopole Cai} and \ref{eq:quadrupole Cai}. Unsurprisingly, this model does not successfully describe the data at any point within the void interior. Outside the mean void radius $\overline{R_v}=55\;h^{-1}$Mpc, the multipoles from this model approach the correct expression: this explains the observation of \citet{Cai:2016a} that the linear growth rate estimator proposed in that paper works only in the region $s>\overline{R_v}$.

Table~\ref{table:chi2} compares the goodness of fit for the various models discussed: here `lin.+dispersion' refers to the full linear dispersion model above, `lin.+dispersion+bias' to this model with the additional simplifying linear bias assumption $\xi^r(r)/b=\delta(r)$, `lin. only' to the model without dispersion, Eq.~\ref{eq:xis full}, and `lin.+bias' to this model with the addition of the bias assumption, Eq.~\ref{eq:xis bias}. The fit for the streaming model of Eq.~\ref{eq:GSM} uses $\delta(r)$ (without the linear bias assumption) to obtain $v_r(r)$, and $\sigma_v(r)$ measured from simulation, so can be directly compared to the dispersion model without bias. The fit for the \citet{Cai:2016a} model includes the bias assumption, as in Eq.~\ref{eq:xis Cai}. 

\section{Measuring the growth rate}
\label{sec:growthrate}

\begin{table}
%\begin{minipage}{0.5\textwidth}
\centering
\caption{$\chi^2/$d.o.f. values for fit to data with different models, using the fiducial growth rate for the simulation. See text for details.}
\begin{tabular}{@{}lccc}
\hline
Model & \multicolumn{3}{c}{$\chi^2/$d.o.f} \\
& monopole & quadrupole & combined \\
\hline
lin.+dispersion & 0.90 & 1.17 & 1.18 \\
\hline
lin.+dispersion+bias & 1.44 & 1.61 & 1.72 \\
lin. only & 5.75 & 5.86 & 7.39 \\
lin.+bias & 8.63 & 8.50 & 10.75 \\
streaming model, Eq. \ref{eq:GSM} & 5.61 & 3.19 & 5.02 \\
\protect\citet{Cai:2016a}, Eq. \ref{eq:xis Cai} & $2.4\times10^4$ & 37.1 & $1.3\times10^4$ \\
\hline\\
\end{tabular}
\label{table:chi2}
%\end{minipage}
\end{table}

%==================Fig.: =======================%
\begin{figure}
\begin{center}
\includegraphics[scale=0.4]{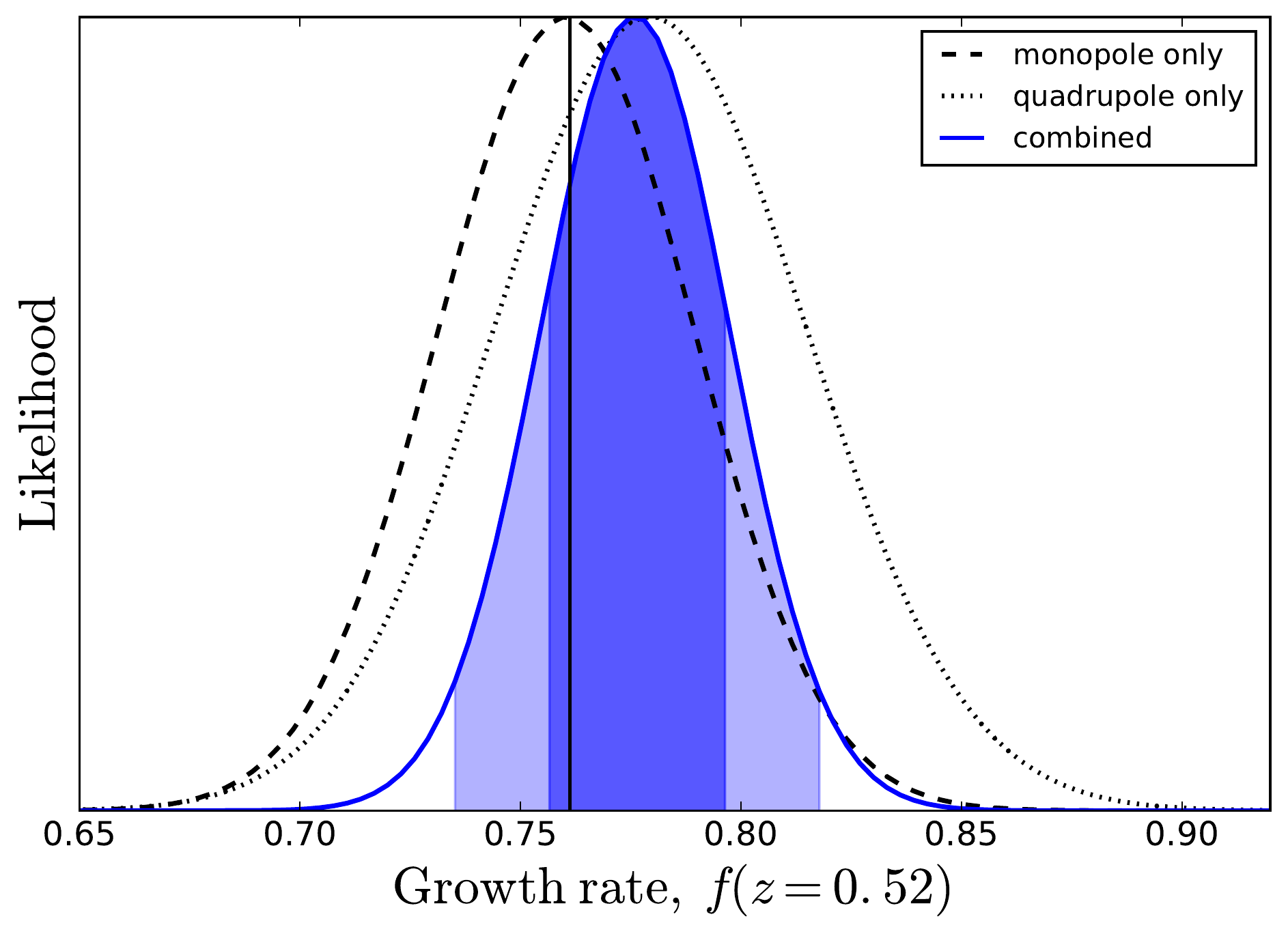}
\caption{Likelihood for the growth rate $f$, obtained from fits of Eq.~\ref{eq:xis dispersion} to the measured multipoles of the void-galaxy cross-correlation, corresponding to a recovered value $f=0.78\pm0.02$, a $2.7\%$ constraint. The dashed and dotted curves show the likelihoods for fits to the monopole and quadrupole separately. The functions $\delta(r)$ and $\sigma_{v_{||}}(r)$ are calibrated from the simulation. The vertical line indicates the fiducial value of the growth rate.}
\label{fig:constraints}
\end{center}
\end{figure}
%==================Fig.:=======================%

We have seen that the fiducial linear dispersion model for the RSD in the void-galaxy correlation provides an excellent description of the multipoles measured from simulation at all scales. An interesting question is how this model might be used to determine the value of growth rate $f$ within voids from fits to data.

As discussed in Section~\ref{sec:multipoles}, it is not possible to construct an estimator for $f$ from the ratio of measured redshift space multipoles. Therefore to fully specify the model and calculate the multipoles for a given value of $f$ in principle requires knowledge of three functions: $\xi^r(r)$, $\delta(r)$, and $\sigma_{v_{||}}^\mathrm{gal}(r)$. All three depend strongly on the void-finding algorithm and the choice of centre for stacking, as well as potentially having a smaller cosmological dependence.

Of these, the most important input to the model is the real space correlation function $\xi^r(r)$, which cannot be measured directly. A plausible procedure would be to calibrate $\xi^r$ using fits to the real space void profile in simulated mocks before use in the RSD model. This is the procedure followed by \citet{Hamaus:2015,Hamaus:2016,Achitouv:2017a,Achitouv:2017b}. Various fitting formulae have been described in the literature, e.g. \citet{Hamaus:2014a,Ricciardelli:2014,Nadathur:2015a,Nadathur:2015b,Nadathur:2015c,Barreira:2015,Cautun:2016}. However, there is no consensus on any particular fitting form, and in any case the fit will strongly depend on the void-finding and stacking algorithms as well as on the properties of the galaxy sample in question \citep{Nadathur:2015c}, so the calibration   needs to be performed on a case-by-case basis. Alternative methods to obtain $\xi^r$ are by deprojecting the redshift space monopole $\xi^s_0$ \citep{Pisani:2014} or by using reconstruction techniques \citep{Nadathur:2018b}.

In principle the matter overdensity profile $\delta^r(r)$ can be determined independently of $\xi^r(r)$ through void lensing measurements \citep{Krause:2013,Melchior:2014,Clampitt:2015,Sanchez:2016}. Alternatively, it could also be calibrated from simulations as for $\xi^r$. Use of the linear bias assumption $\delta(r)=\xi^r(r)/b$ is however \emph{not} a good approximation if the void-galaxy correlation is to be determined for the same galaxy tracers used to identify the voids. This is because of the selection effect described in Section \ref{sec:bias}, that introduces a systematic scale-dependent shift to the inferred relationship between $\xi^r(r)$ and $\delta(r)$. For instance, using the linear bias approximation for our simulation data leads to errors in the predicted quadrupole $\xi^s_2$ of up to $20\%$.

The other piece of information required to perform a fit for $f$ is the dispersion relation $\sigma_{v_{||}}(r)$. It is possible that future work on a complete streaming model for the void-galaxy case may allow this function to be determined theoretically from $\xi^r(r)$ and $\delta(r)$, or it could be calibrated from simulation in the same way as $\delta(r)$. At present we take its value directly from the simulation measurements. This leaves only a single free parameter in our model: the growth rate $f$.

Allowing $f$ to vary and fitting to the measured multipole data using Eq.~\ref{eq:xis dispersion}, we obtain the posterior likelihoods shown in Figure~\ref{fig:constraints}. This gives a recovered value of $f=0.78\pm0.02$ ($68\%$ c.l.), consistent with fiducial value $f=0.761$ for the simulation, and corresponding to a $2.7\%$ precision for our simulation volume of $(2.5\;h^{-1}\mathrm{Gpc})^3$. The constraints obtained from the monopole and quadrupole taken individually are also shown, and are consistent with each other and with the combined result.

Table~\ref{table:rcuts} shows how the constraints on $f$ change when only data within different separation ranges are used for the fits. In all cases the constraints obtained are consistent with each other and the final value. It is also clear that almost all of the constraining power of the data comes from the contribution to $\xi^s$ from galaxies within the void interiors, $s<\overline{R_v}$, which is the region where our model performs significantly better than the alternatives.

Finally, we tested the effect on the measurement of the growth rate of using the linear bias assumption $\delta(r)=\xi^r(r)/b$. Due to the failures described above, this leads to a strongly biased estimator of $\beta$, with the reconstructed value being more than $3\sigma$ smaller than the fiducial for fits to both the quadrupole and monopole.

A popular alternative methodology \citep[e.g.][]{Hamaus:2015,Cai:2016a,Achitouv:2017b} is not to fix $\sigma_{v_{||}}(r)$ from simulation but to parametrise it using some functional form with one or more free parameters, which are to be marginalised over. The simplest possible such parametrisation is to take it to be a constant,
\begin{equation}
\label{eq:sigma_v}
\sigma_{v_{||}}(r)=\sigma_0.
\end{equation}
Figure~\ref{fig:velocity} suggests that this might be a reasonable approximation separately in the void interior and exterior regions. As sensitivity to the dispersion also decreases in the void exterior, where $\sigma_v\ll raH$, it might be hoped that this approximation is sufficient to reconstruct $f$.

%==================Fig.: =======================%
\begin{figure}
\begin{center}
\includegraphics[scale=0.4]{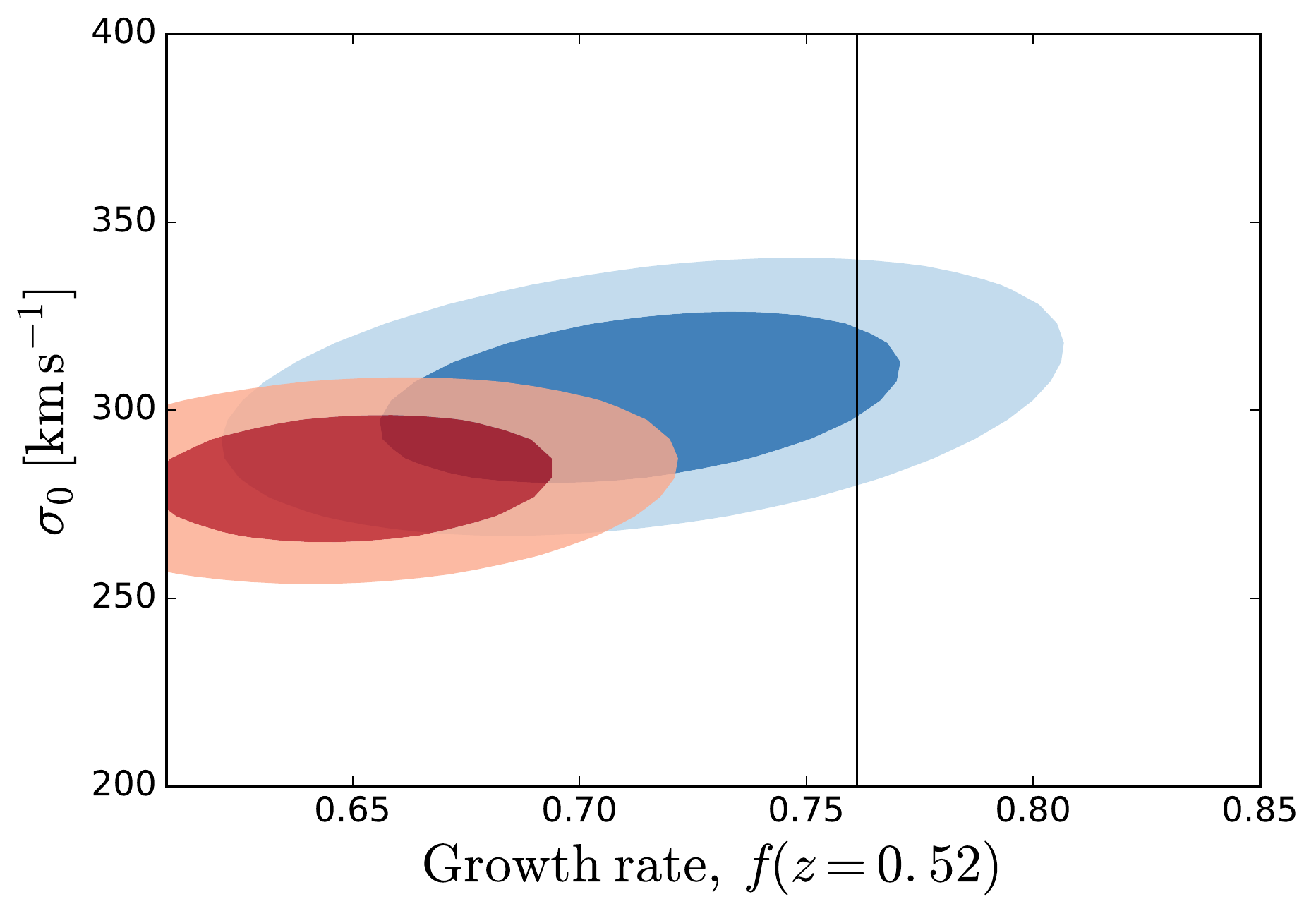}
\caption{$68\%$ and $95\%$ confidence limit contours on the two free parameters $\left(f,\sigma_0\right)$, where $\sigma_0$ is a constant velocity dispersion, from fits to the quadrupole (blue contours) and monopole (red) from simulation data. The vertical line shows the fiducial value of $f$ for the simulation. Introducing $\sigma_0$ as a free parameter weakens the constraints on $f$. The quadrupole still provides an unbiased estimate of $f$, with $5.3\%$ precision, but the monopole estimator is biased.}
\label{fig:2Dconstraints}
\end{center}
\end{figure}
%==================Fig.:=======================%

\begin{table}
%\begin{minipage}{0.5\textwidth}
\centering
\caption{Constraints on $f$ obtained from fits to data in different separation ranges relative to the mean void radius $\overline{R_v}=55.6\;h^{-1}\mathrm{Mpc}$.}
\begin{tabular}{@{}lc}
\hline
Data range & $f$ \\
\hline
$s<0.5\overline{R_v}$\ & $0.79\pm0.03$ \\
$s<\overline{R_v}$\ & $0.77\pm0.02$ \\
%$s>0.5\overline{R_v}$\ & $0.74\pm0.03$ \\
%$s>\overline{R_v}$\ & $0.82\pm0.05$ \\
all $s<120\;h^{-1}\mathrm{Mpc}$\ & $0.78\pm0.02$ \\
\hline\\
\end{tabular}
\label{table:rcuts}
%\end{minipage}
\end{table}

To test this, we fit the model to the measured multipoles from the simulation data with $f$ and $\sigma_0$ as two free parameters. $\xi^r(r)$ and $\delta(r)$ are taken from the fits to the simulation data as before. Figure~\ref{fig:2Dconstraints} shows the resulting $1$ and $2\sigma$ confidence level contours determined from fitting the quadrupole (blue) and monopole (red) separately. It is clear that the addition of the additional parameter loosens the constraints obtained considerably. The fit to the quadrupole provides an unbiased reconstruction of the fiducial growth rate, with the reconstructed value $f=0.72\pm0.04$ at $68\%$ c.l. after marginalizing over $\sigma_0$, corresponding to a $5.3\%$ constraint lower than but consistent with the fiducial value $f=0.761$. However, fitting to the monopole $\xi^s_0$ gives a biased reconstruction of the growth rate which is more than $3\sigma$ from the fiducial value. This indicates a failure of the constant dispersion model and limits the amount of information that can be extracted from measurement of $\xi^s$ if a constant $\sigma_0$ is assumed.

\section{Conclusions}
\label{sec:conclusions}

Measurement of redshift space distortions in the void-galaxy correlation function $\xi^s(\mathbf{s})$ is an important tool that can be used to test for possible environmental dependence of the growth rate of structures. In particular, the growth rate in the lowest density regions close to void centres might contain information about possible non-standard theories of gravity. However, reconstructing the growth rate in voids requires a model for $\xi^s$ which can be trusted in these regions.

We have derived a configuration-space model for the void-galaxy correlation in redshift space using linear theory, which we characterise in terms of the multipole moments of the correlation. Our model accounts for several terms that are important within voids but have previously been neglected. As a result we are able to account for important physical effects that had not been appreciated, including the change in sign of the quadrupole term within the void, indicating a turnover point between stretching and squashing of the contours of the correlation function along the line of sight direction. The model can be broadened to include a dispersion in galaxy velocities along the line of sight; the dispersion model thus obtained differs from the streaming model used in previous studies, which was based on an inappropriate application of the formula for the Gaussian streaming model derived for the galaxy autocorrelation.

Comparing our model predictions to measurements of $\xi^s$ using void and galaxy catalogues at redshift $z=0.52$ in the Big MultiDark simulation shows that the linear dispersion model provides an excellent fit to the data for \emph{all} values of the void-galaxy separation down to the minimum bin width used, $2.4\;h^{-1}$Mpc. This is an important contrast to the modelling of RSD in the galaxy correlation, where non-linear effects are very important at small pair separations and complicate the use of small-scale data. Comparison with the data shows that our linear model for the void-galaxy correlation performs significantly better than others in the literature, especially within the low density region of interest. 

A consequence of our results is that the ratio of the redshift space monopole and quadrupole does not factorize to give an estimator for the growth rate as for the model of Eq. \ref{eq:xis Cai}. However, as discussed above, this model does not fit the data within the void interior, at $s<\overline{R_v}$. The quadrupole-to-monopole ratio has previously been used in other works \citep{Cai:2016a,Hamaus:2017a}; however, \citet{Nadathur:2018b} show that if used outside its regime of validity it can lead to strongly biased estimates of the growth rate.

Determination of the growth rate using the correct expression we provide requires knowledge of the real space correlation function $\xi^r(r)$, which can be taken from fits to simulations or possibly reconstructed from redshift space data. Using functional forms for $\delta(r)$ and $\sigma_{v_{||}}(r)$ determined from the simulation data, we show that fitting for the growth rate yields $f=0.78\pm0.02$, a $2.7\%$ constraint in good agreement with the fiducial value $f=0.761$. Assuming a constant dispersion with amplitude $\sigma_{v_{||}}(r)=\sigma_0$ taken as a free parameter, fits to the quadrupole alone still provide an unbiased estimate of the true growth rate $f$ after marginalising over $\sigma_0$, though with a reduced precision of $5.3\%$. However, the assumption of a constant dispersion leads to a biased reconstruction of the true $f$ from the monopole data. 

Our results also highlight two very important points which have relevance to practical applications of this method to measure RSD effects within voids. Firstly, if the galaxy tracers used to measure the RSD effects are the same as those used to identify the voids, an assumption of linear galaxy bias within the void interior is not appropriate due to statistical selection effects on the void size scale. Assuming a constant linear bias leads to incorrect predictions from the model, which possibly require the addition of empirical non-linear correction terms \citep{Achitouv:2017b}, even though in reality linear theory continues to provide a good description of the dynamics. In order to overcome this problem, in practical terms it would be better to use two differently biased galaxy tracers covering the same volume to identify void regions and to measure the RSD effects.

Secondly, we highlighted the assumptions inherent in the derivation -- in particular, the conservation of void numbers and the invariance of void positions under the redshift space mapping -- that are required to hold for consistent comparison with the data. These assumptions are not unique to the model of the void-galaxy RSD presented in this paper, but are key features of \emph{all} models in the literature \citep{Paz:2013,Cai:2016a,Achitouv:2017a,Achitouv:2017b}. They hold by construction in our simulation data, because we identify voids in the real space galaxy field. In general if the voids are identified using galaxies in redshift space, these assumptions fail \citep{Zhao:2016,Chuang:2017,Nadathur:2018b}. 

In survey data, where only redshift space information is available, a practical method to reconstruct the real space void positions in order to allow fair comparison of theory and data for $\xi^s$ has been described by \citet{Nadathur:2018b}. This method is based on using a reconstruction algorithm similar to that used for BAO detection in order to undo RSD effects and approximately recover the real-space galaxy distribution before performing void-finding. \citet{Nadathur:2018b} show that the voids thus obtained match the true real-space void population closely, thus satisfying the requirements for the modelling presented in this work. This demonstration is for the same galaxy mocks as used in this work, which are designed to match those in the SDSS BOSS data releases. The success of reconstruction depends on the galaxy density, and is not expected to work in the limit of poor sampling; however, note that successful reconstruction is only required at galaxy locations as the shifts are applied to galaxy positions. Reconstruction is also expected to fail in very high density regions due to multistreaming, but for voids we are generally not interested in such regions. These considerations mean that the reconstruction method of \citet{Nadathur:2018b} is expected to be applicable for all large-volume spectroscopic galaxy redshift surveys such as BOSS, eBOSS, DESI or Euclid, where the same technique is also used for BAO detection. That is, these surveys are designed to allow reconstruction to improve BAO measurements, and where this is possible the same process can always be used to obtain unbiased estimates of the growth rate from void-galaxy RSD, as described by \citet{Nadathur:2018b}.
 
Finally, a comment on the relative merits of using the void-galaxy correlation and the galaxy autocorrelation to measure RSD is in order. It is true that unlike for the galaxy RSD case, the model presented in this work is based only on linear theory, and fits the simulation data extremely well at all pair separations. It is particularly noteworthy that dispersion effects can be easily accommodated within a linear model. These appear to be significant advantages over the galaxy RSD case, for which quasi-linear modelling is required and small-scale data is excluded when performing fits. However, since most of the information in the void-galaxy correlation comes from those galaxies in the void interiors, this method also discards some of the available data. One therefore should not necessarily expect it to outperform the standard analysis, as has sometimes been suggested \citep{Hamaus:2017a}. Instead an advantage of the void-galaxy RSD measurement is specifically to constrain possible environmental dependence of the growth rate in low-density regions. It also provides a complementary measurement technique that is sensitive to different systematics. 

The model presented in this work depends only on the growth rate $f$, with functions $\delta(r)$ and $\sigma_{v_{||}}(r)$ calibrated directly from simulation for a single cosmological model. In practical application of this technique to analysis of data from redshift surveys, a number of additional parameters would be introduced, which could be treated as nuisance parameters and marginalised over. These include $\sigma_8$ (which would modulate the calibrated amplitude of $\delta(r)$), bias $b$ (which is required for reconstruction; \citealt{Nadathur:2018b}), and $\sigma_{v_{||}}$.

\section{Acknowledgements}

We thank Yanchuan Cai for helpful correspondence, and Davide Bianchi, Florian Beutler, Hans Winther, Rossana Ruggeri and Eva-Maria M\"uller for stimulating discussions and comments. SN acknowledges funding from the Marie Sk\l odowska-Curie Actions under the H2020 Framework of the European Commission, project 660053 {\small COSMOVOID}. WJP acknowledges support from the European Research Council through the Darksurvey grant 614030, and from the UK Science and Technology Facilities Council grant ST/N000668/1 and the UK Space Agency grant ST/N00180X/1. The BigMultiDark simulations were performed on the SuperMUC supercomputer at the LeibnizRechenzentrum in Munich, using resources awarded to PRACE project number 2012060963. Numerical computations were performed on the {\small EREBOS}, {\small THEIA} and {\small GERAS} clusters at the Leibniz Institut f\"ur Astrophysik (AIP).

\bibliographystyle{mnras}
\bibliography{refs.bib}

\appendix
\section{Effects of void centre definition}
\label{appendix:centres}

%==================Fig.: =======================%
\begin{figure*}
\begin{center}
\includegraphics[scale=0.5]{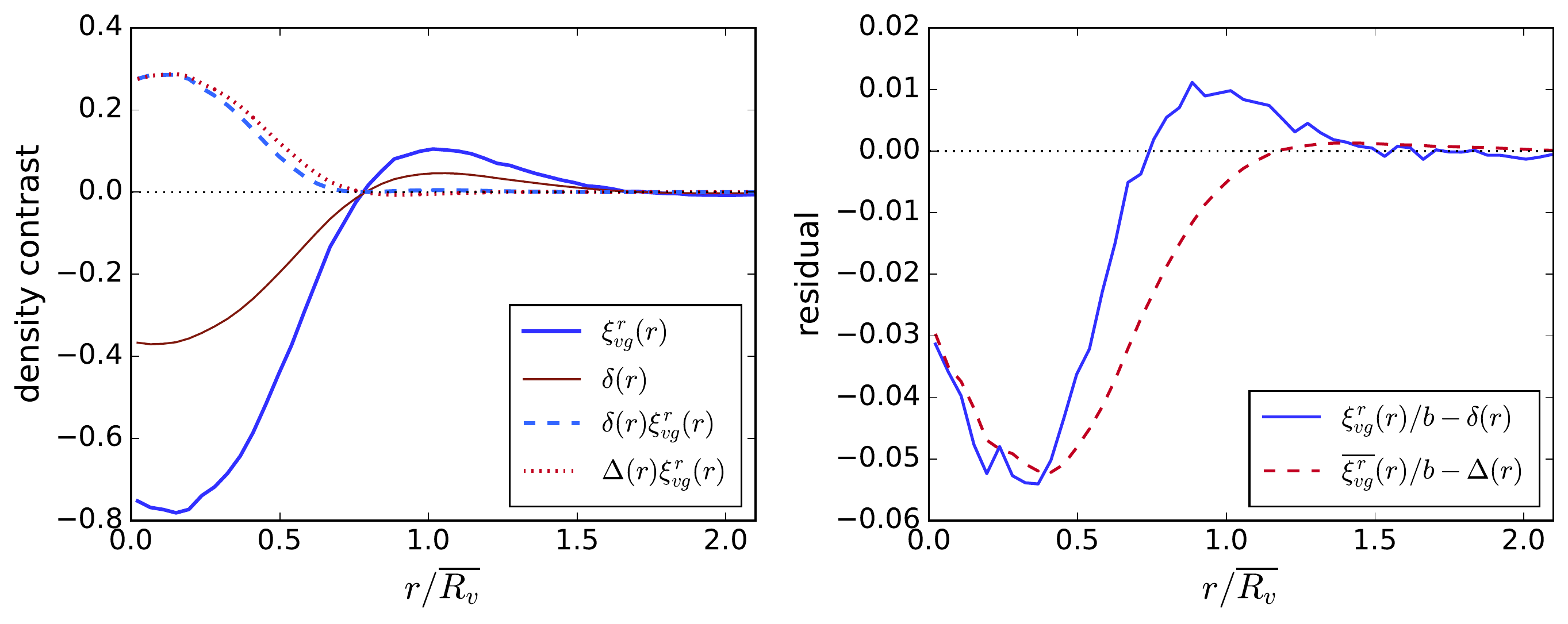}
\caption{Same as Figure~\ref{fig:approximations}, and for the same void and galaxy data, but with distances from the void centres defined relative to the void barycentres.}
\label{fig:approximations_baryC}
\end{center}
\end{figure*}
%==================Fig.:=======================%

%==================Fig.: =======================%
\begin{figure*}
\begin{center}
\includegraphics[scale=0.35]{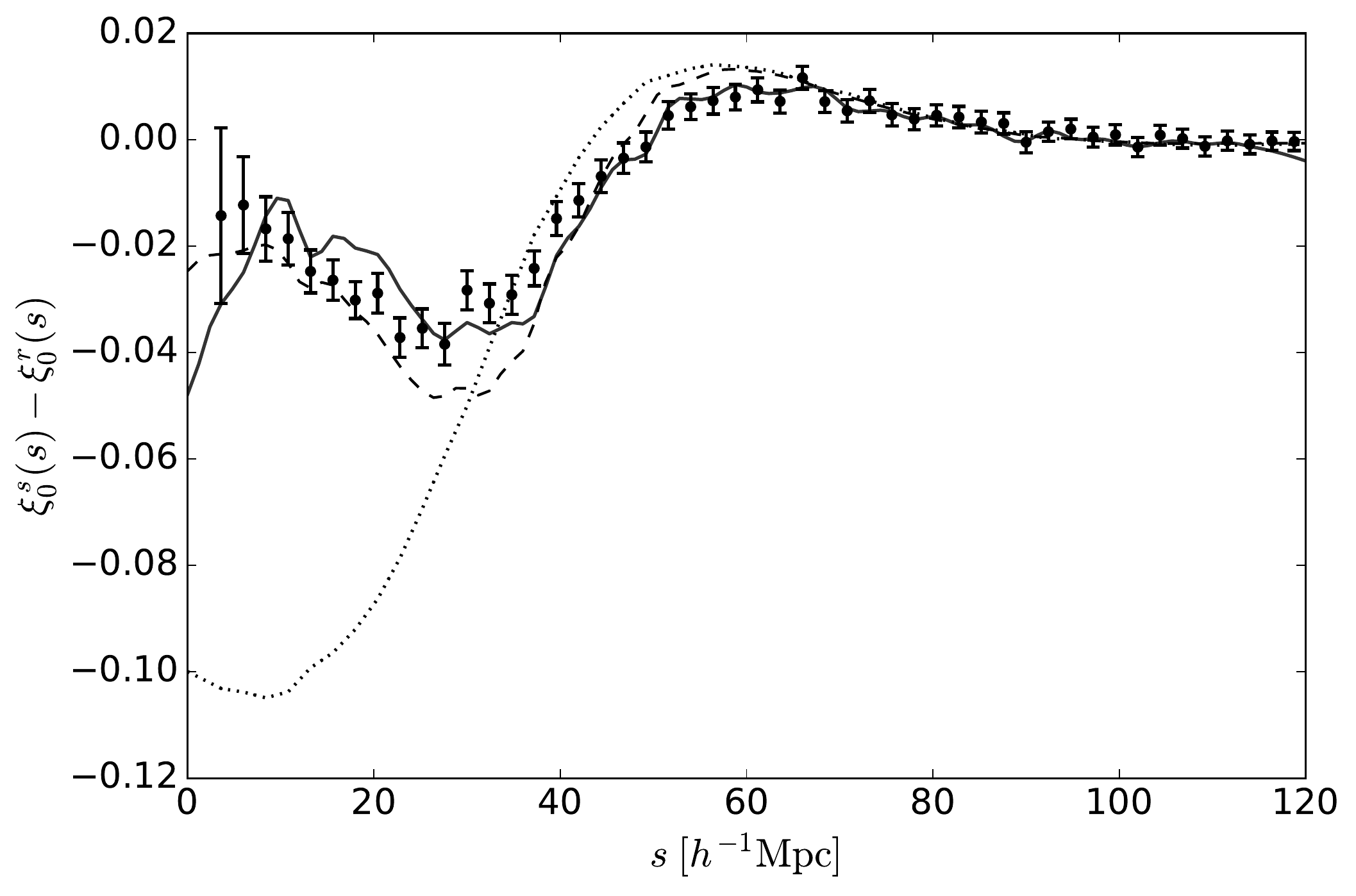}
\includegraphics[scale=0.35]{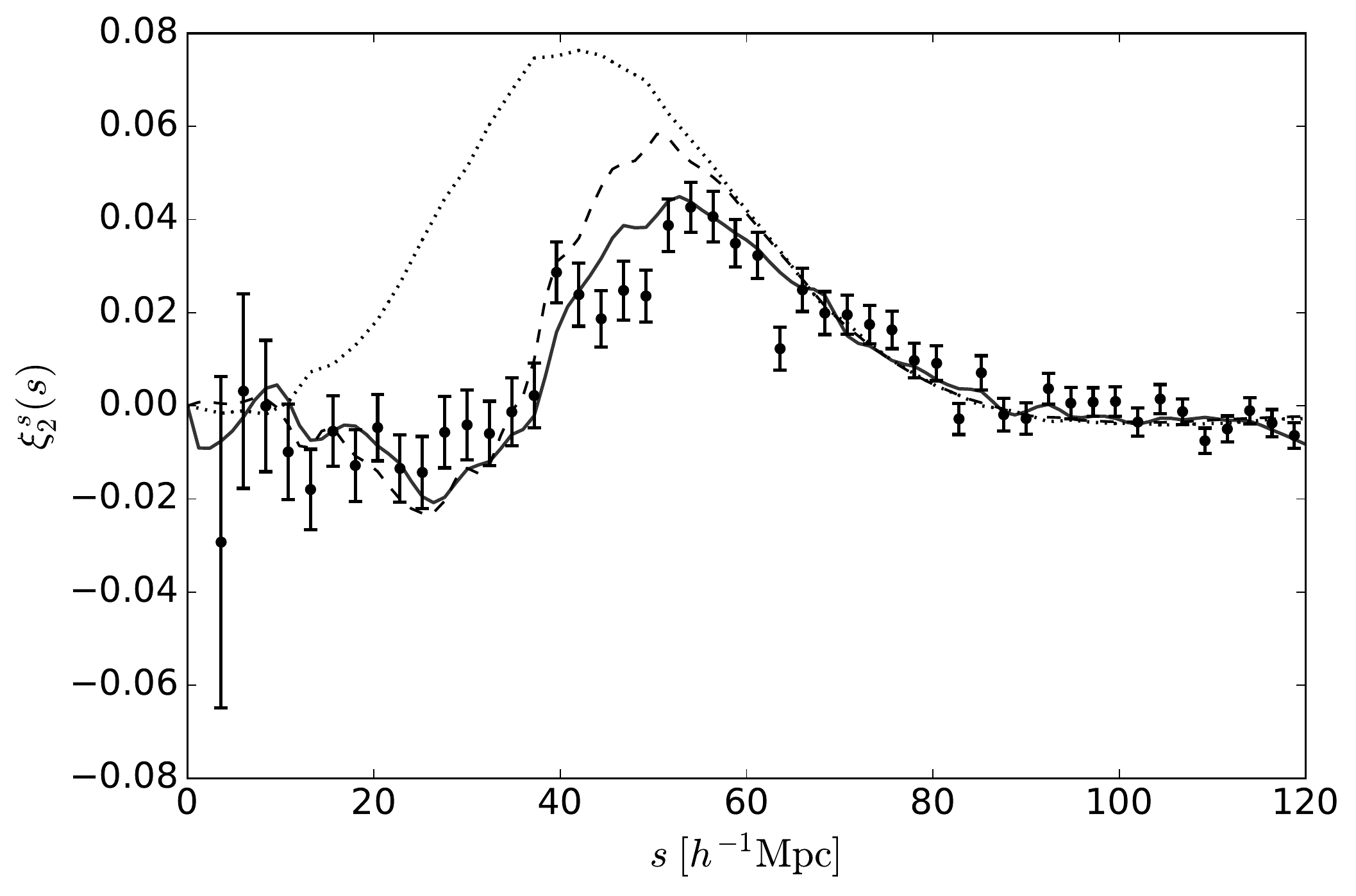}
\caption{Same as Figure~\ref{fig:model multipoles}, but for stacks centred on the void barycentre positions.}
\label{fig:multipoles baryC}
\end{center}
\end{figure*}
%==================Fig.:=======================%

Throughout this work, we have used the void centre definition introduced by \citet{Nadathur:2015b}, which places the void centre at the point furthest removed from the galaxy positions, i.e. the point of minimum galaxy density within the void. The choice of void centre is however not unambiguous, and alternative definitions have previously been used in the literature. In particular, \citet{Hamaus:2017a} define the void centre to be the volume-weighted barycentre of the positions of galaxies within the void. The choice of centre for stacking can affect details of the measured $\xi_{vg}$ values in both real and redshift space. However the derivation of the theoretical model makes no assumptions about the choice of centre. The purpose of this appendix is to explicitly demonstrate that that the primary conclusions of this work, viz.:
\begin{enumerate}
\item the terms proportional to $\xi\delta$ and $\xi\Delta$ are comparable to $\delta$ and $\Delta$, and must be retained in the expansion of Eq.~\ref{eq:xis invJ};
\item the linear bias relationship $\xi^r(r)=b\delta(r)$ does not apply within voids;
\item in the void interior regions, the model proposed in this paper, Eq.~\ref{eq:xis dispersion}, performs significantly better than that of Eq.~\ref{eq:xis Cai},
\end{enumerate}
are independent of the choice of void centre.

The first two of these statements are illustrated by Figure~\ref{fig:approximations_baryC}, which shows the same profiles and residuals as in Figure~\ref{fig:approximations} except with distances measured relative to the void barycentres. The left panel shows that the measured $\xi^r(r)$ and $\delta(r)$ profiles differ somewhat from those in Figure~\ref{fig:approximations}, in particular showing lower density contrasts at the void centres. This is expected, because the barycentre position is known to be a worse tracer of the region of underdensity within a void \citep{Nadathur:2015b,Nadathur:2015c,Nadathur:2017a}. In particular, $\xi^r(r)$ shows a characteristic increase at small $r$ as noted in several previous works. However, the terms $\delta(r)\xi^r(r)$ and $\Delta(r)\xi^r(r)$ are still clearly of the same order as $\delta(r)$ within the void interior. The right panel shows that the linear bias approximation also does not apply for barycentre stacks, with a similar pattern of large and scale-dependent residuals from the bias relationship being observed. Neglecting both these effects, as done by \citet{Cai:2016a} and \citet{Hamaus:2017a} in deriving Eq.~\ref{eq:xis Cai}, therefore still results in an incomplete model independent of the choice of centre.

To demonstrate this explicitly, Figure~\ref{fig:multipoles baryC} shows the multipoles $\xi^s_0(s)$ and $\xi^s_2(s)$ measured from the simulation for the same voids as in Figure~\ref{fig:model multipoles}, but for stacks based around the void barycentres. The curves shown on the plot correspond to the same theoretical models as before, i.e. Eq.~\ref{eq:xis dispersion} (solid line), Eq.~\ref{eq:xis full} (dashed), and Eq.~\ref{eq:xis Cai} (dotted). The theoretical curves differ from those shown in Figure~\ref{fig:model multipoles} because the input $\xi^r(r)$, $\delta(r)$ etc. differ for barycentre stacks.

%==================Fig.: =======================%
\begin{figure}
\begin{center}
\includegraphics[scale=0.4]{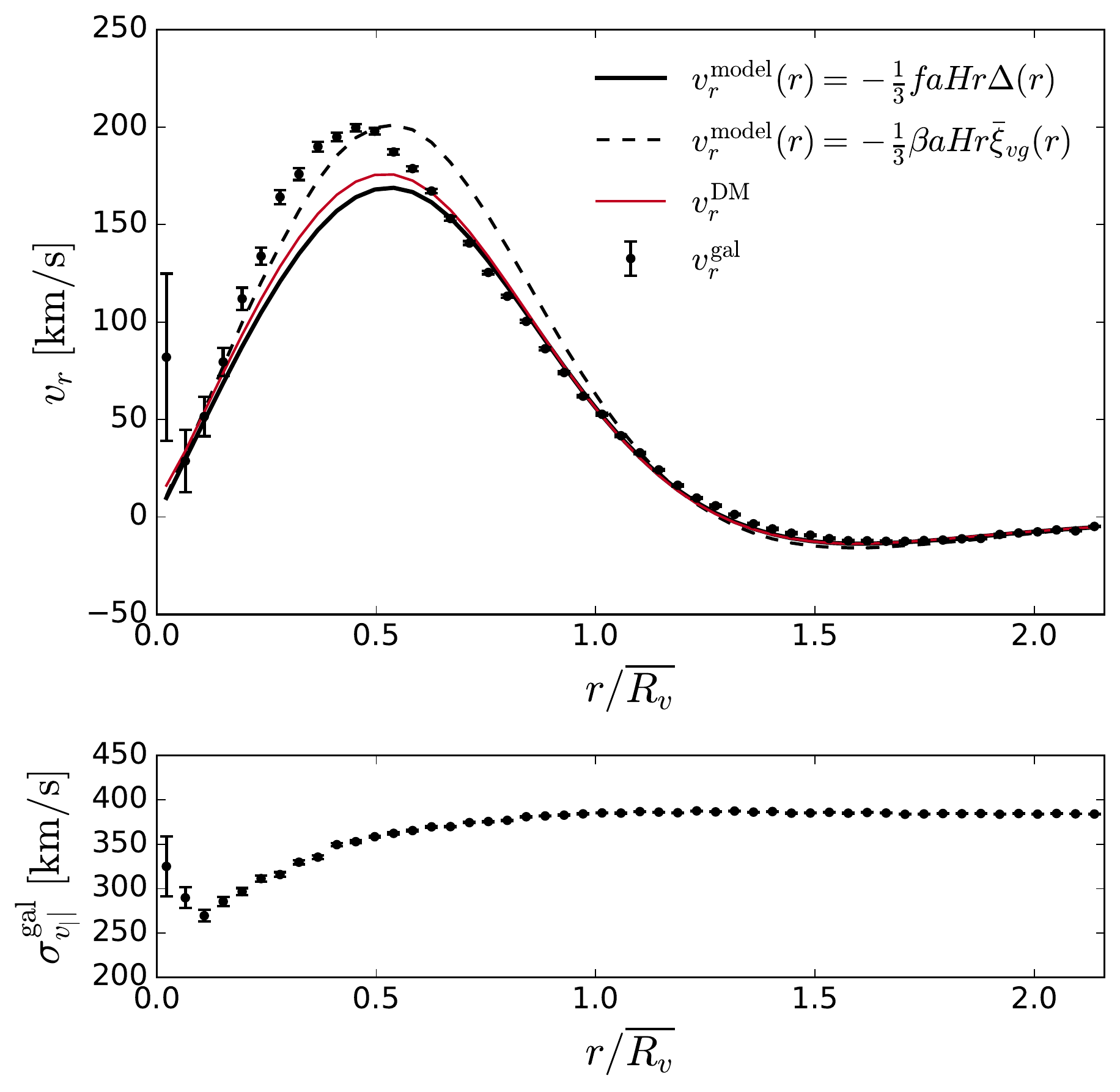}
\caption{Same as Figure~\ref{fig:velocity}, but for stacks centred on the void barycentre positions.}
\label{fig:velocity baryC}
\end{center}
\end{figure}
%==================Fig.:=======================%

Comparison of theory and data shows the same qualitative conclusions hold as before: i.e., Eq.~\ref{eq:xis Cai} does not correctly describe the measured multipoles, whereas Eq.~\ref{eq:xis dispersion} provides a much better description. However, Figures \ref{fig:model multipoles} and \ref{fig:multipoles baryC} also reveal import differences. Firstly, for both the monopole and the quadrupole, the amplitude of the RSD effect is smaller for the barycentre stacks, and the data is significantly noisier, with much larger error bars, as we have also checked by comparison of the full covariance matrices. This means that stacking around the void barycentre is sub-optimal for void RSD measurements. Secondly, the linear dispersion model of Eq.~\ref{eq:xis dispersion}, while qualitatively still correctly capturing the physics, now provides a significantly worse quantitative fit to the data: the reduced $\chi^2$ values are now 1.63, 1.37 and 1.76 for fits to monopole, quadrupole, and both combined. These should be compared to the first line of Table~\ref{table:chi2}.

The reason for this deterioration of the model fit can be understood by reference to Figure~\ref{fig:velocity baryC}. This shows that for barycentre stacks, the measured velocity profiles $v_r^\mathrm{gal}(r)$ (data points) and $v_r^\mathrm{DM}(r)$ (thin solid curve) both differ much more strongly from the linear theory prediction based on Eq.~\ref{eq:linear velocity} (thick solid curve) at small $r$ than for the stacks around the minimum density centre (Figure~\ref{fig:velocity}). This means that the assumption of the validity of the linear dynamics governed by the void density profile $\Delta(r)$ alone is less valid for stacks around the barycentre. A likely reason for this is that since void barycentre positions do not trace the position of the true minimum density within the void, they also are worse tracers of the stationary points of the velocity field; that is, large-scale velocity gradients across the void leading to deviations from Eq.~\ref{eq:linear velocity} are more significant for stacks centred at the void barycentre positions. This explanation is consistent with the results of \citet{Nadathur:2017a}, who show that void barycentre positions are worse tracers of maxima of the gravitational potential, where $\nabla\Phi=0$ and thus the linear velocity outflow is locally spherically symmetric.

\label{lastpage}

\end{document}